\documentclass[aps,prd,twocolumn,noeprint,showpacs,nofootinbib,superscriptaddress]{revtex4-1}

\usepackage[colorlinks=true,citecolor=blue,linkcolor=blue,breaklinks=true]{hyperref}
\usepackage{graphicx}
\usepackage{color}
\usepackage{natbib}
\usepackage{epsfig}
\usepackage{float}
\usepackage{xcolor}
\usepackage{url}
\usepackage{bm}

\usepackage{amsmath,amssymb,amsfonts}

\newcommand{\aap}{Astron. Astrophys.}%
\newcommand{\prx}{Phys. Rev. X}

\begin{document}

\title{Compact Dark Objects in Neutron Star Mergers}

\author{Andreas Bauswein}
\affiliation{GSI Helmholtzzentrum f\"ur Schwerionenforschung, Planckstra{\ss}e 1, 64291 Darmstadt, Germany}
\affiliation{Helmholtz Research Academy Hesse for FAIR (HFHF), Campus Darmstadt, 64291 Darmstadt, Germany}

\author{Gang Guo}
\affiliation{Institute of Physics, Academia Sinica, Taipei, 11529, Taiwan}
\affiliation{School of Mathematics and Physics, China University of Geosciences, Wuhan 430074, China}

\author{Jr-Hua Lien}
\affiliation{Institute of Physics, Academia Sinica, Taipei, 11529, Taiwan}

\author{Yen-Hsun Lin}
\affiliation{Institute of Physics, Academia Sinica, Taipei, 11529, Taiwan}
\affiliation{Physics Division, National Center for Theoretical Sciences, Taipei 10617, Taiwan}

\author{Meng-Ru Wu}
\affiliation{Institute of Physics, Academia Sinica, Taipei, 11529, Taiwan}
\affiliation{Institute of Astronomy and Astrophysics, Academia Sinica, Taipei, 10617, Taiwan}
\affiliation{Physics Division, National Center for Theoretical Sciences, Taipei 10617, Taiwan}

\date{\today}


\begin{abstract} 
We estimate the long-lasting gravitational wave (GW) emission of compact dark objects following a binary neutron-star (NS) merger. We consider compact dark objects, which initially reside in the centers of NSs and which may consist of self-interacting dark matter (DM). By approximating the compact dark objects as test particles, we model the merging of NS binaries hosting DM components with three-dimensional relativistic simulations. 
Our simulation results suggest that the DM components remain gravitationally bound and orbit inside the merger remnant with orbital separations of typically a few km. 
The subsequent orbital motion of the DM components generates a GW signal with frequencies in the range of a few kHz. 
When considering a range of different binary masses and high-density equations of state (EoS), we find that the GW frequency of the orbiting DM components scales with the compactness of NSs.
Similarly, we find relations between the DM GW frequency and the dominant postmerger GW frequency of the stellar fluid or the tidal deformability, which quantifies EoS effects during the binary inspiral. 
Hence, a measurement of these quantities can be used to specify the frequency range of the GW emission by DM. 
Under the assumption that GW back reaction is the only relevant dissipative process, the GW signal may last between seconds and years depending on the mass of the DM component. 
We estimate the detectibility of the GW signals and find that DM components in NS mergers may only be detectable with existing and projected GW instruments if the dark objects are as massive as about $0.01$ to $0.1~M_\odot$. 
We emphasize that the GW emission is limited by the lifetime of the remnant. A forming black hole will immediately swallow the DM objects because their orbits are smaller than the innermost stable circular orbit of the black hole.
\end{abstract}



\maketitle   

\section{Introduction}


Concurrent searches for particle physics candidates of dark matter (DM) via
direct and indirect methods have been pursued extensively over the last decades (see, e.g., Refs.~\cite{Gaskins:2016cha,Kahlhoefer:2017dnp,Buchmueller:2017qhf,Boveia:2018yeb,Liu:2017drf,Schumann:2019eaa,PerezdelosHeros:2020qyt,PDG2020} for recent reviews).
Among these efforts, the potential existence and effects of DM inside neutron stars (NSs) were widely explored, including DM-triggered gravitational collapse \cite{Kouvaris:2010jy,Kouvaris:2011gb,Bramante:2013nma,Garani:2018kkd,Bell:2018pkk,Lin:2020zmm,Dasgupta:2020dik} as well as heating resulting from
DM annihilation \cite{Kouvaris:2007ay,deLavallaz:2010wp,Chen:2018ohx,Lin:2021hen} and scattering off baryons and/or leptons \cite{Baryakhtar:2017dbj,Raj:2017wrv,Acevedo:2019agu,Camargo:2019wou,Joglekar:2020liw,Bell:2020jou,Bell:2020lmm}.
Motivated by the prospects in the era of the gravitational wave (GW) astronomy, several studies discuss how current and future GW observations can probe the nature of DM (for a review, see e.g.,~Ref.~\cite{Cardoso:2019rvt}). These include the effect of DM on the tidal deformability of NSs, which can be measured in binary NS merger events, e.g., GW170817~\cite{Nelson:2018xtr,Ellis:2018bkr,Das:2018frc,Quddus:2019ghy,Das:2020vng,Ivanytskyi:2019wxd,Zhang:2020pfh,Dengler2022,Cassing:2022tnn,Leung:2022wcf,2023arXiv230304089F}. 
Additional signatures may occur via long-range forces in the dark sector during the inspiral phase of NS mergers~\cite{Kopp:2018jom,Alexander:2018qzg,Choi:2018axi}, the postmerger GW emission~\cite{Ellis:2017jgp,Bezares:2019jcb}, GW signals by dark object orbiting inside galactic NSs~\cite{Horowitz:2019aim}, and effects on the X-ray pulse profile~\cite{Miao2022ApJ...936...69M}.

Regarding the postmerger GW emission produced with DM-admixed NSs, Ref.~\cite{Ellis:2017jgp} adopted a theory-agnostic approach using a ``mechanical model'', which relies on a number of free parameters to estimate the GW spectrum within the first $\sim 10$~ms after merging. 
It was found that DM cores of $\simeq 0.1~M_\odot$ inside NSs can potentially produce additional peaks in the postmerger GW spectrum with similar strength as those produced by the baryonic components. 
On the other hand, Ref.~\cite{Bezares:2019jcb} took a different approach by performing general relativistic hydrodynamic simulations of equal-mass binaries with two 1.35~$M_\odot$ NSs, each containing a bosonic DM core with a mass fraction of 5\% or 10\%. The chosen parameters were such that the bosonic DM core had a size comparable to the fermionic part of the NS.
Based on their setup, the authors found that for cases with a 10\% DM core, a one-arm instability can develop on a $\sim \mathcal{O}(100)$~ms timescale, which damps the dominating quadrupolar GW emission from the baryonic component. See also~\cite{Rafiei2022,Giangrandi2022} for the consideration of bosonic DM.

The effects of mirror dark matter in NS mergers has been studied within a two-fluid treatment in~\cite{Emma2022}. See also~\cite{Ruter2023} for two-fluid binary NS initial data.

In this work, we consider the postmerger GW emission of DM-admixed binary NS mergers within a scenario different from the aforementioned studies~\cite{Ellis:2017jgp,Bezares:2019jcb}. 
Motivated by the work of Ref.~\cite{Horowitz:2019aim}, we explore the possibility of long-term postmerger GW emission \emph{predominantly} from orbiting DM cores. We consider dark compact objects with sizes $\lesssim 1$~km, thus much smaller than typical NS radii, and with masses in the range of  $\lesssim 10^{-2}$~$M_\odot$.
One example for dark compact objects of this kind may be models having a self-interacting, asymmetric fermion DM candidate as proposed in, e.g., a series of work by Refs.~\cite{Gresham:2017cvl,Gresham:2017zqi,Gresham:2018anj,Gresham:2018rqo}.
In particular, Ref.~\cite{Gresham:2018rqo} showed that such a highly-compact dark object can be stable when the strong self-interaction is balanced by gravity (for self-repulsive force) or by the enhanced degenerate pressure due to the reduced DM particle mass at high densities (for self-attractive force). 
See Appendix~\ref{app:dark_radius} for details.
Note that for a DM-admixed NS of this type, the structure of the baryonic matter outside the dark core is barely influenced by the presence of the light core (see Ref.~\cite{Gresham:2018rqo}), and can be well-described by the solution of a pure baryonic NS.

The formation channels of asymmetric DM-admixed NSs is yet under intensive investigations.
The most studied mechanism is via the capture of halo DM through their scattering with baryons~\cite{Kouvaris:2007ay,Guver:2012ba,McDermott:2011jp,Garani:2018kkd,Bell:2020jou}, which, however, can only accumulate up to $\lesssim 10^{-13}M_\odot$ of DM.
Other scenarios include the production of DM during the supernova explosions~\cite{Nelson:2018xtr}, via the anomalous decay channel of neutrons~\cite{McKeen:2018xwc,Cline:2018ami,Grinstein:2018ptl}, or through the combination of the anomalous neutron decay and the capture of DM~\cite{Keung:2020teb}.
All these scenarios can possibly result in NSs with DM cores of sub-solar masses with small coupling between the DM and nucleons below the current experimental reaches.

Relevant to the evolution of the DM cores in the NS postmerger remnants is whether one should consider their interaction with baryons. 
For a given DM--nucleon cross-section $\sigma_{\chi N}$ and a characteristic relative velocity of order $\sim 0.1$~c in the merger remnants, the minimal interaction timescale per dark matter particle with ambient hadronic matter is $\sim \mathcal{O}(1)$~s~$\times (10^{-48}{\rm cm}^2/\sigma_{\chi N})$ for $\rho_N\simeq 5\times 10^{14}$~g~cm$^{-3}$ without taking into account the further suppression of Pauli-blocking. 
The Pauli-blocking effect of both degenerate nucleons and DM can significantly reduce the scattering rates by at least several orders of magnitudes\footnote{The timescale for DM-baryon interaction can be estimated by $\tau^{-1}_{\rm int}\sim n_b \sigma_{\chi n} v_{\rm rel} \xi$ where $\xi$ is the Pauli blocking suppression factor. 
For DM thermalization inside a NS,
previous estimates showed that $\xi \ll 10^{-5}$ for non-degenerate DM~\cite{Bertoni:2013bsa}. 
For highly-degenerate DM in a compact dark object of our concern, this will further suppress the interaction between DM and baryons and completely justify to ignore the DM--baryon collisions.}.

Furthermore, for DM cores lighter than $\lesssim 10^{-2}$~$M_\odot$, one may neglect the gravitational force, which the DM component exerts on the baryonic matter.
Based on all the above estimates, we take a simplified approach by approximating DM cores inside the NSs as \emph{test particles}. We follow their trajectories with relativistic hydrodynamic NS merger simulations out to $\sim 30$~ms after collision when the merger remnants become approximately axisymmetric and quasi-stationary. We describe the dynamics in Sec.~\ref{sec:simulation}.
In Sec.~\ref{sec:results}, we investigate the dependence on nuclear physics and on the binary system using four different nuclear EoSs and four mass configurations, and analyze the simulation results.
In Sec.~\ref{sec:detectability}, we follow a similar methodology as outlined in Ref.~\cite{Horowitz:2019aim} to estimate the long-term GW emission for different DM core masses and evaluate the detectibility with current and next-generation GW telescopes. 
We discuss potential caveats and conclude in Sec.~\ref{sec:conclusion}.

In this study we set $G=c=1$ unless we explicitly provide units. Latin indices run from 1 to 3, greek indices run from 0 to 3. Stellar masses refer to the gravitational mass in isolation, in binary systems masses are given at infinite orbital separation. We define the binary mass ratio as $q=M_1/M_2$ with $M_1\leq M_2$. Neutron star radii refer to the circumferential eigen radius.

\section{Simulations}\label{sec:simulation}

{\it Numerical tool:} We perform relativistic hydrodynamical simulations of NS mergers with a relativistic smoothed particle hydrodynamics (SPH) code, which has been successfully used in various previous studies~\cite{Oechslin2002,Oechslin2007,Bauswein2012a,Bauswein2019b}. The code employs the conformal flatness condition on the spatial metric $\gamma_{ij}=\psi^4\delta_{ij}$ to solve the Einstein field equations within the 3+1 split of space time~\cite{Isenberg1980,Wilson1996,Baumgarte2010}. The hydrodynamical equations are closed by the EoS, which is provided in the form of temperature and composition dependent tables.

In addition to the hydrodynamical evolution, we evolve the motion of test particles in the gravitational field by implementing the geodesic equation in the code. Here we explicitly assume that on the relevant time scales DM only interacts through gravity and that the contribution of DM to the gravitational field is negligible. The latter certainly holds for not too large DM masses. We estimate that this model may yield quantitatively reliable results for DM masses below $\sim 0.01~M_\odot$. Already for DM masses as low as $0.01~M_\odot$ we expect a significant phase shift during the inspiral phase comparing systems of the same baryon mass with and without DM\footnote{Neutron stars harboring substantial amounts of compact DM will lead to more compact stellar configurations as compared to stars of the same total mass without DM. Hence, we speculate that finite-size effects during the GW inspiral will be affected by the presence of DM (see also~\cite{Ellis:2018bkr,Das:2018frc,Quddus:2019ghy}). Binary events with the same masses but different tidal deformability may be an indicator of such a scenario. Considering that only DM masses above $\sim0.01~M_\odot$ lead to radius modifications in excess of 100~m, we suspect that the effect may be generally difficult to measure and to discern from other uncertainties during the inspiral.} (as a result of a secular effect during the long lasting evolution). We suspect that the angular momentum and density distribution at the end of the inspiral is very similar in both systems and thus the postmerger evolution should be well captured by our model.

In the 3+1 split the geodesic equation becomes
\begin{equation}
    \frac{du_i}{dt}=-\alpha u^0\partial_i\alpha+u_k\partial_i\beta^k-\frac{1}{2 u^0}u_j u_m \partial_i\gamma^{jm}
\end{equation}
with the coordinate velocity
\begin{equation}
    \frac{dx^j}{dt}=\frac{u^j}{u^0}=\gamma^{jk}\frac{u_k}{u^0}-\beta^j
\end{equation}
and the Lorentz factor being $\alpha u^0=\sqrt{1+\gamma^{ij}u_iu_j}$ (see e.g.~\cite{Baumgarte2010}). Here $\alpha$ refers to the lapse function, $\beta^i$ is the shift vector and $u^\mu$ denotes the eigen velocity. Within the conformal flatness approximation these equations become
\begin{equation}
\begin{split}
    &\frac{du_i}{dt}= -\alpha u^0\partial_i\alpha+u_k\partial_i\beta^k +\frac{2u_ku_k}{\psi^5u^0}\partial_i\psi\\
     &\frac{dx^j}{dt}=\frac{u_j}{\psi^4 u^0}-\beta^j.
\end{split}    
\end{equation}
The very same equation can be directly obtained from the Lagrangian formulation of the relativistic hydrodynamical equations by setting the pressure $P=0$ and the specific enthalpy $h=1$ in the momentum equation (cf.~\cite{Siegler2000,Oechslin2007}). In essence, we obtain a set of ordinary differential equations describing the motion of test particles with strong similarities to the hydrodynamical equations in relativistic SPH. Exploiting this similarity we implement the equations for the motion of DM with a Runge-Kutta method, which is solved in parallel to the evolution of the conserved hydrodynamical variables employing the existing numerical infrastructure.

{\it Initial data and setup:} The simulations start from quasi-stationary orbits a few revolutions before merging. For this we assume zero-temperature stellar matter in neutrinoless beta-equilibrium and we set the intrinsic spin of the stars to zero. In this initial setup we place a test particle in the centers of each binary component. Hence, we assume that DM settled to the center of each star during the long inspiral phase (typically expected to last several 100 million years) and that DM closely follows the motion of the stars. Considering the free evolution of DM particles during the simulated last orbits before merging (see below) fully confirms the latter assumption.

We emphasize that we describe the entire DM contribution by only two test particles originally placed in the center of the stars. Hence, we implicitly assume that one can neglect the contribution to gravity and the spatial extension of these DM components, i.e. treat them as point particles. Because of these choices, we do not need to specify masses of the DM components a priori but can simply adopt DM masses for computing for instance the GW signal, which obviously represents a good approximation as long as the DM mass is sufficiently small. Our approximation implies another caveat namely that point particles do not collide, whereas finite size self-interacting DM components can touch and interact, which cannot be captured by our calculations.

\begin{figure*}
\centering
\includegraphics[width=0.98\columnwidth]{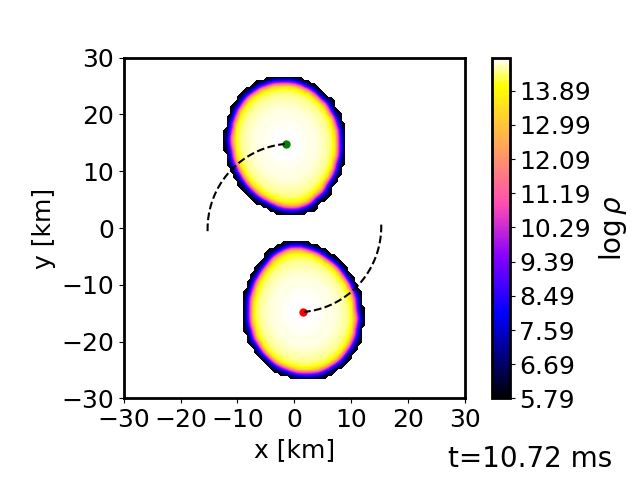} 
\includegraphics[width=0.98\columnwidth]{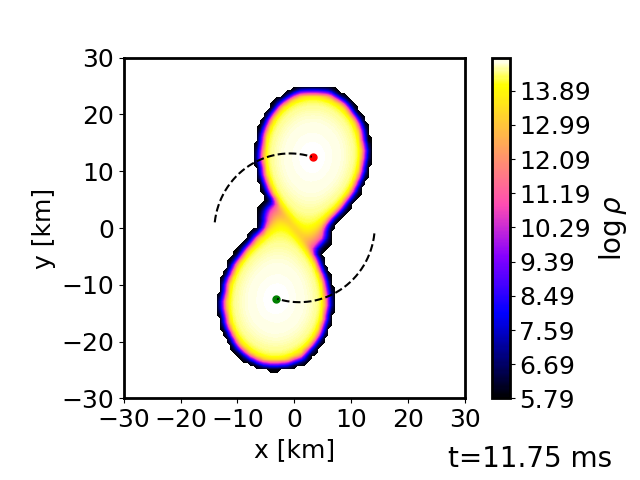} \\
\includegraphics[width=0.98\columnwidth]{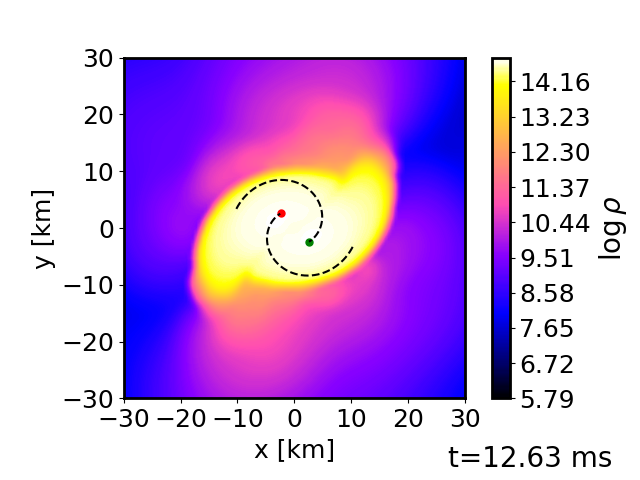} 
\includegraphics[width=0.98\columnwidth]{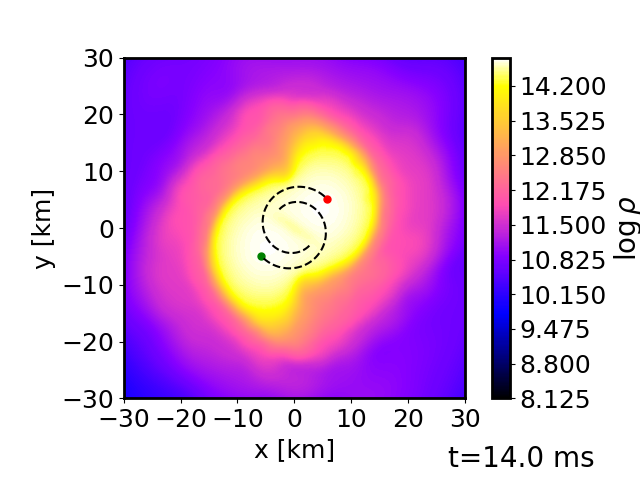} \\
\includegraphics[width=0.98\columnwidth]{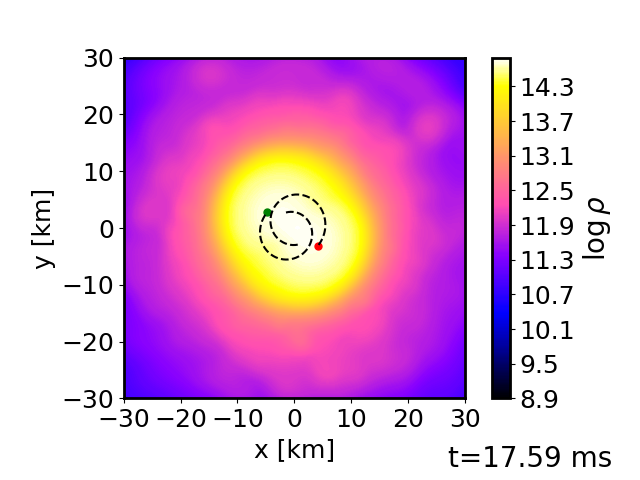} 
\includegraphics[width=0.98\columnwidth]{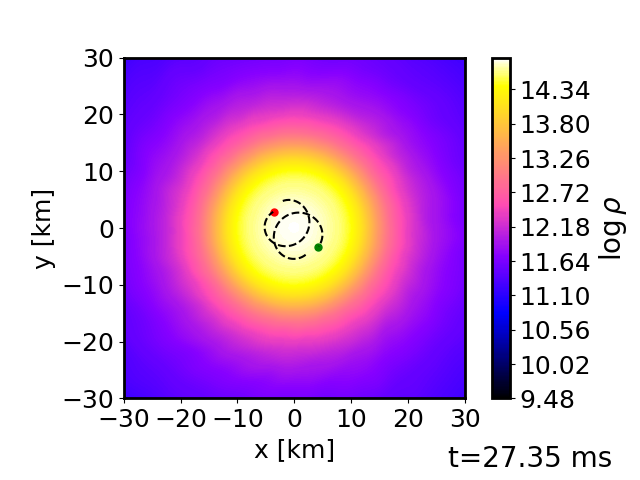} 

\caption{Rest-mass density in the equatorial plane (color-coded; $\rho$ in $\mathrm{g/cm^3}$) for a 1.35-1.35~$M_\odot$ merger with the DD2 EoS. Green and red dots display the position of test particles. The dashed lines indicate the trajectories of the test particles during the last 0.5~ms w.r.t. the given time step. (The unsmooth edges at the surface of the stars are a result of the visualization routine.)}
\label{fig:snap}
\end{figure*}

In this study we consider three different binary systems with a total binary mass of $M_\mathrm{tot}=2.7~M_\odot$ and binary mass ratios of $q=M_1/M_2=\{1,1.279/1.421,1.2/1.5\}$. In addition, we simulate equal-mass mergers with $M_\mathrm{tot}=2.4~M_\odot$. To assess the impact of the EoS of NS matter and to generally test the robustness of our results, we employ four different temperature-dependent models of nuclear matter for each of the aforementioned binary systems. We use the SFHO, SFHX, DD2 and DD2F EoSs~\cite{Typel2005,Hempel2010,Typel2010,Steiner2013,Alvarez-Castillo2016}, which are compatible with current astrophysical constraints on the maximum NS mass, tidal deformability and radius~\cite{Antoniadis2013,Cromartie2019,Abbott2019}. These models roughly span a considerable range of NS properties.

{\it Simulations:} All binary configurations which we consider in this study lead to the formation of a NS postmerger remnant, while binaries with higher total mass may directly form a BH sensitively depending on the EoS~\cite{Bauswein2020}. Figure~\ref{fig:snap} shows the evolution of a 1.35-1.35~$M_\odot$ binary merger in the equatorial plane for the DD2 being representative for the other simulated systems. Before merging the stars exhibit significant tidal deformations. The early postmerger remnant is strongly deformed and oscillates (see also GW spectrum in Fig.~\ref{fig:spec}). On a time scale of several 10~ms the central object approaches axisymmetry and quasi-stationarity. During the secular evolution angular momentum is redistributed and the central density of the remnant slowly increases. A delayed gravitational collapse of the merger remnant can occur either during the early dynamical phase or the secular evolution. (The specific model in Fig.~\ref{fig:snap} does not undergo a collapse until the end of the simulation. Considering the relatively high maximum mass of nonrotating NSs for the employed EoS, delayed BH formation may only take place on a very long time scale if magnetic dipole spin down triggers the collapse of the finally rigidly rotating remnant.) Obviously, the results of these simulations are fully in line with previous merger calculations since we explicitly neglect the back reaction of DM on the merger dynamics.

In Fig.~\ref{fig:snap} we display the location of the two DM test particles by a green and a red dot. The dashed lines visualize the trajectory of the respective particle during the last 0.5~ms with respect to the shown simulation snapshot. The two upper panels clearly show that during the inspiral the DM particles closely follow the center of mass of the respective binary components. The DM remains confined to the local gravitational potential well. During merging the DM particles are injected on orbits in the now forming common potential well of the merger remnant and decouple from the fluid motion. The test particles remain gravitationally bound to the system and orbit around each other, while the gravitational field of the remnant undergoes a dynamical evolution in this early phase. We refer to~\cite{Bauswein2016} for plots of the lapse function in the orbital plane, which indicates the gravitational potential. (In the weak field limit of Einstein's equations the lapse function is related to the Newtonian gravitational field.)

\begin{figure}
\centering
\includegraphics[width=0.98\columnwidth]{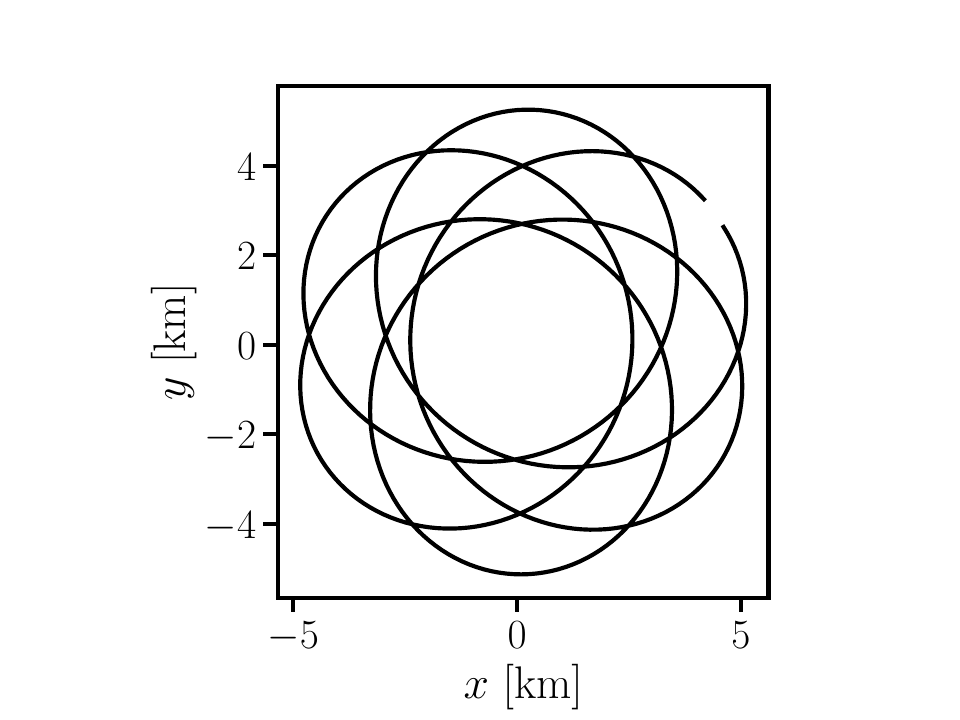} 
\caption{Trajectory of DM component during last 5~ms of the simulation of a 1.35-1.35~$M_\odot$ binary with the DD2 EoS.}
\label{fig:orb}
\end{figure}

\begin{figure}
\centering
\includegraphics[width=0.98\columnwidth]{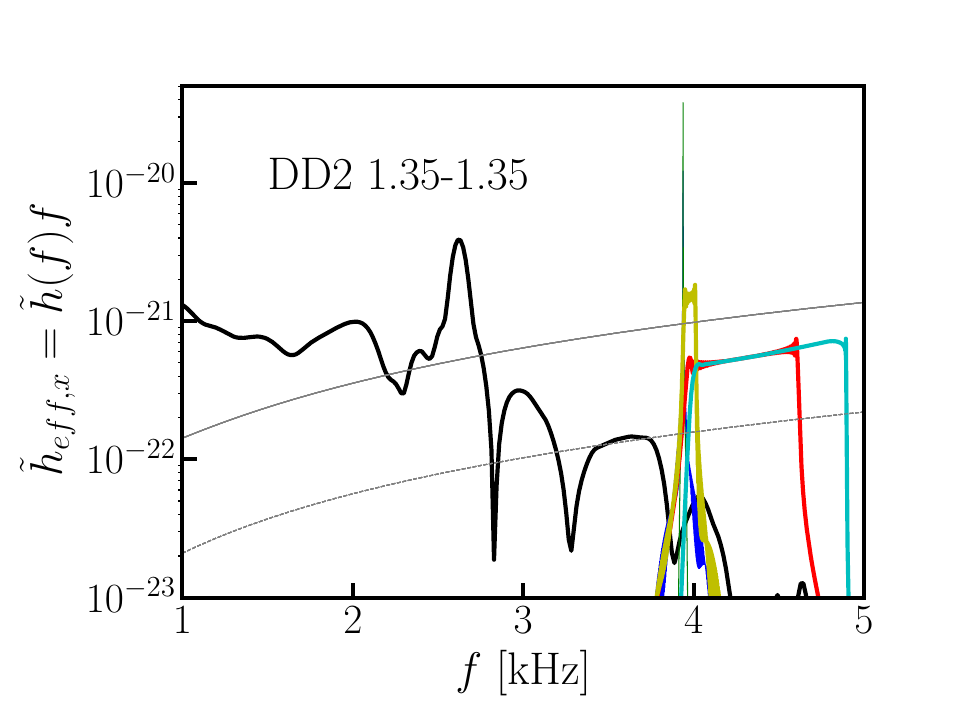} 
\caption{Gravitational-wave spectrum of the cross polarization for a 1.35-1.35~$M_\odot$ merger with the DD2 EoS at a  distance of 20~Mpc (black). Colored curves display GW emission by DM component under different assumptions about lifetime and density gradient inside the central remnant: blue=\{1~s,~0\%\}, green=\{10~s,~0\%\}, red=\{1~s,~200\%\}, yellow=\{1~s,~10\%\}, and cyan=\{10~s,~100\%\} assuming a DM mass of 0.01~$M_\odot$ per component (see text for details and in particular Appendix~\ref{app:longterm} for the simplistic description of a density gradient inside the remnant). Blue and green curves overlap in the figure. Gray curves show unity-SNR sensitivity for Advanced Ligo (solid) and the Einstein Telescope (dashed). Considering a density gradient, peaks broaden and become flatter but SNRs do not change significantly. See also Appendix~\ref{app:noncircular} for a discussion of the impact on non-circular orbits of the DM particles.}
\label{fig:spec}
\end{figure}

As a result of the merger dynamics and the subsequent remnant evolution, the orbits of the DM particles are elliptic and show a strong periastron advance. As an example we plot the trajectory of one DM component extracted from the 1.35-1.35~$M_\odot$ simulation with the DD2 EoS. Figure~\ref{fig:orb} visualizes the DM motion during the last five ms of the simulation. Within our model we neglect any interaction between DM and the fluid, which means that the DM particles can potentially orbit for a very long time inside the quasi-stationary remnant. Hence, already small DM masses may generate a considerable GW signal if the emission continues over a longer period (cf.~\cite{Horowitz:2019aim}). By treating DM as test particles during the relatively short simulation time we ignore the back reaction of GW emission, which will lead to a slow orbital decay and decrease of the orbital eccentricity on a longer time scale. On time scales beyond the simulation time, the GW back reaction has to be taken into account.

We provide a quantitative analysis of the orbits in the next section discussing the resulting GW emission of DM and the secular decay of the orbits by continued GW emission. We already stress that the orbits in the different simulations are generally relatively tight with orbital radii of only a few km. This implies that these orbits and especially their periastron are smaller than the innermost stable circular orbit of a BH which may form in a delayed collapse of the remnant. Hence, the DM particles would be quickly swallowed by the BH. This implies that GW emission of DM particles is ultimately limited by the lifetime of the merger remnant. Therefore, binary systems with lower masses may generally provide better prospects to find GW emission from DM, although one may speculate that lighter NSs may contain less DM, which in turn slightly decreases the GW amplitude. In this context we also refer to one additional simulation of a binary with two stars of 1.5~$M_\odot$ using the SFHO EoS, which for this EoS leads to the prompt formation of a BH. We find that the BH immediately accretes the DM components.

In Fig.~\ref{fig:spec} we show the GW spectrum of the merger simulation described above, i.e. a 1.35-1.35~$M_\odot$ binary with the DD2 EoS. The black curve displays the signal produced by the baryonic matter. The dominant postmerger oscillation in this model generates a pronounced peak at about $f_\mathrm{peak}=2.6$~kHz. The power below $\sim 1.5$~kHz results from the very last inspiral phase before merging. Note that apart from the main peak several other oscillations modes produced subdominant features below and above $f_\mathrm{peak}$~\cite{Bauswein2019b}. As will be discussed later, those subdominant high-frequency peaks can be comparable in strength and frequency to potential GW emission from DM particles (colored curves in Fig.~\ref{fig:spec}). However, there is little risk to confuse the different emission mechanisms because orbiting DM particles generate a weak, long-lasting signal, which may only become detectable for a sufficiently long remnant lifetime and integration time. The oscillations of the stellar fluid cease within a few 10~ms. Similarly, the oscillations of a forming BH during a prompt or delayed collapse emit GWs at frequencies of several kHz but on relatively short time scales. The NS remnant may possibly be subject to the Chandrasekhar-Friedman-Schutz (CFS) instability and emit GW at later times (see e.g.~\cite{Doneva2015}). This effect is not captured by our model. (Note that the model of DM employed in~\cite{Bezares:2019jcb} also results in a relatively short emission of GWs). Obviously, the quantitative details of the GW spectrum of the baryonic matter sensitively depend on the EoS and the binary masses~\cite{Bauswein2019b}. The dominant oscillation typically occurs at a frequency between 2 and 4~kHz. We finally remark that because of the signal length very different GW data analysis strategies may be required for the oscillations of stellar matter and the emission of DM. With regard to the latter one may employ techniques for periodic sources while a frequency and amplitude evolution should be expected as we discuss below.

\section{Results: DM frequencies and orbits}\label{sec:results}
We analyze the trajectories of DM particles in our simulations to understand their GW signature, which we discuss in more detail in the next section. Already at the end of the simulations at about a few 10~ms after merging, the stellar fluid settles to a quasi-stationary state undergoing a secular evolution with for instance a slow increase of the central densities. As a result of the quasi-stationary gravitational field, the DM particles exhibit a regular orbital motion on elliptic orbits around the center of the remnant including a sizeable periastron advance (see Fig.~\ref{fig:orb}).

\begin{figure}
\centering
\includegraphics[width=\columnwidth]{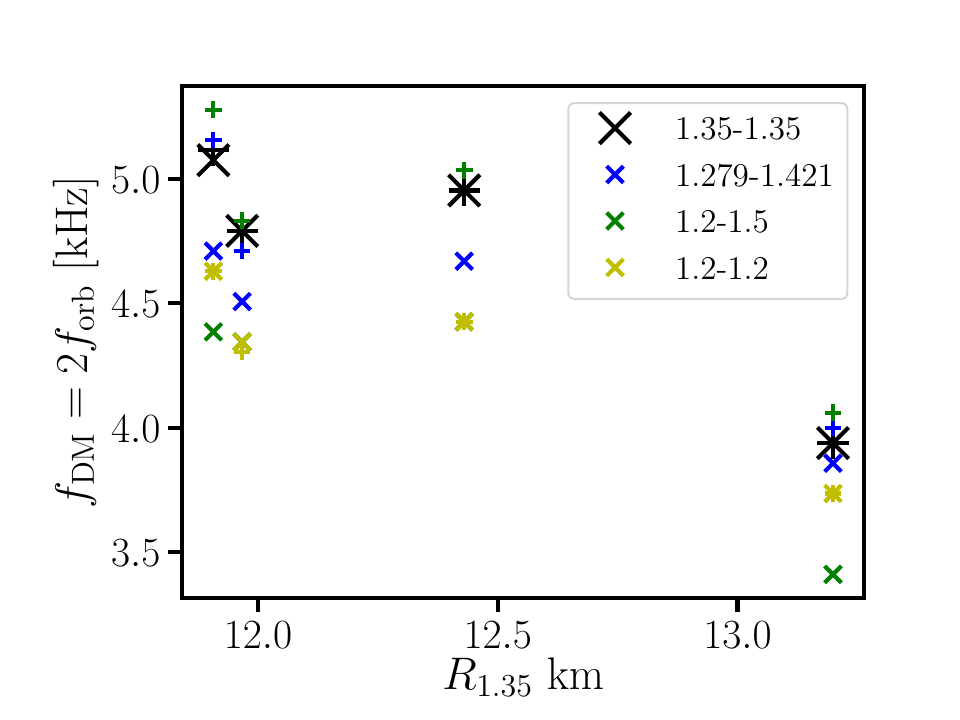}
\caption{GW frequency of DM particles corresponding to twice the orbital frequency as function of NS radius. Colors indicate different binary systems specified in the legend. Per binary systems two data points are given: for asymmetric binaries the plus sign displays DM from the more massive binary component, while the cross shows the test particle stemming from the less massive star. Note that some symbols overlap (e.g. blue and green plus at 12.4~km, and the green and yellow cross at 12.4~km).}
\label{fig:fr135}
\end{figure}

\begin{figure}
\centering
\includegraphics[width=\columnwidth]{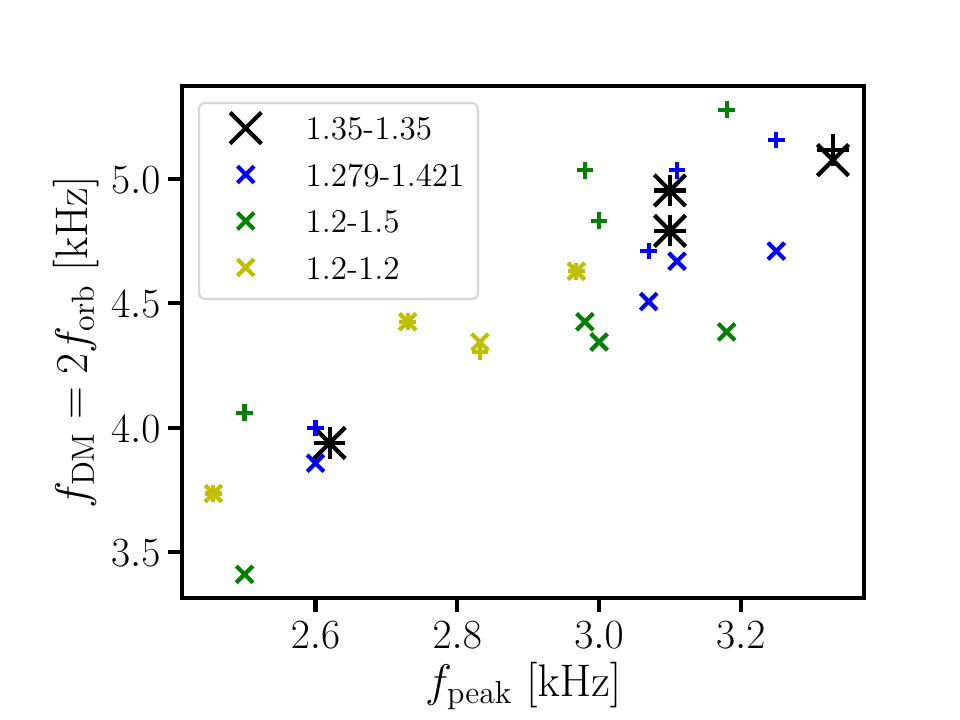}
\caption{GW frequency of DM particles corresponding to twice the orbital frequency as function of dominant postmerger GW frequency. Symbols have the same meaning as in~Fig.~\ref{fig:fr135}.}
\label{fig:fpeak}
\end{figure}

\begin{figure}
\centering
\includegraphics[width=\columnwidth]{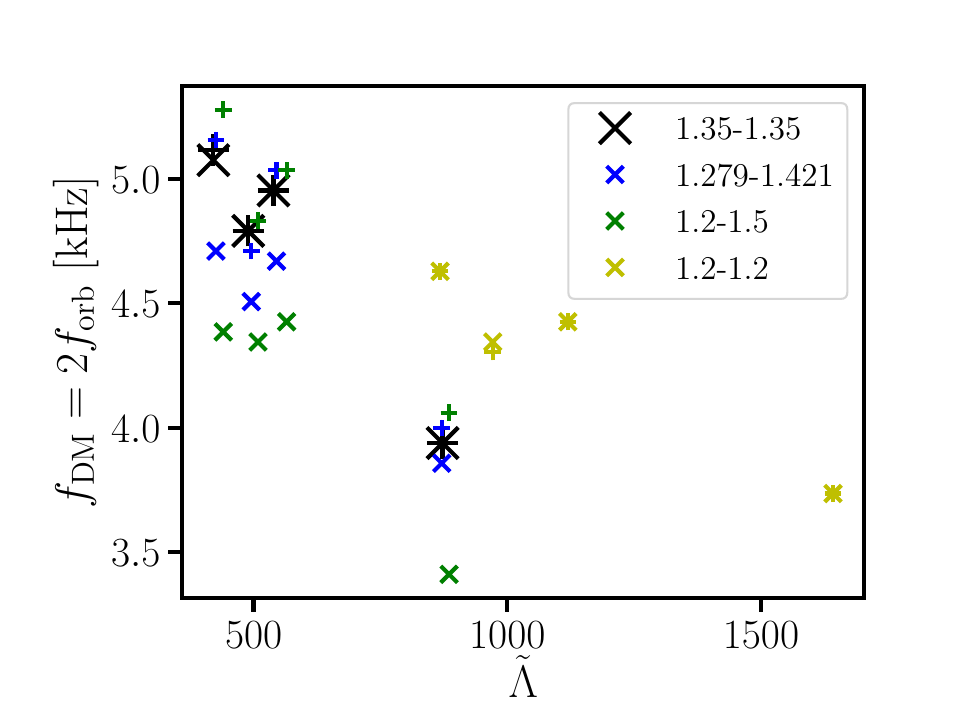}
\caption{GW frequency of DM particles corresponding to twice the orbital frequency as function of combined tidal deformability.  Symbols have the same meaning as in~Fig.~\ref{fig:fr135}.}
\label{fig:lam}
\end{figure}

\begin{figure}
\centering
\includegraphics[width=\columnwidth]{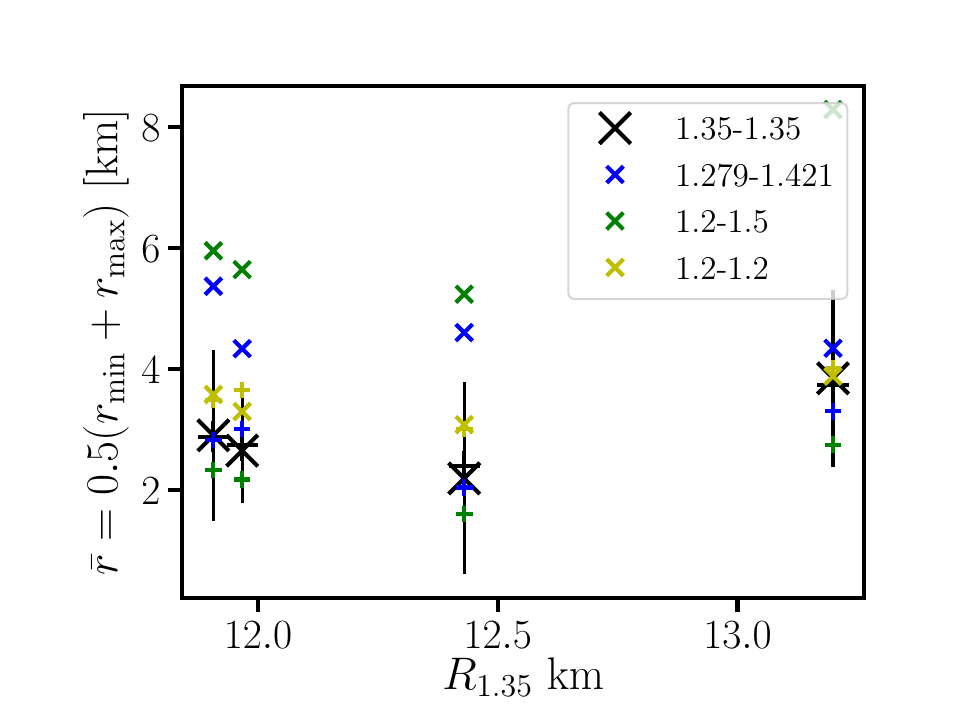}
\caption{Averaged orbital radius of DM particles at the end of the simulations (a few 10~ms after merging) as function of NS radius. Symbols have the same meaning as in~Fig.~\ref{fig:fr135}. Error bars indicate for the equal-mass mergers the minimum and maximum orbit.}
\label{fig:rad}
\end{figure}

\begin{figure}
\centering
\includegraphics[width=\columnwidth]{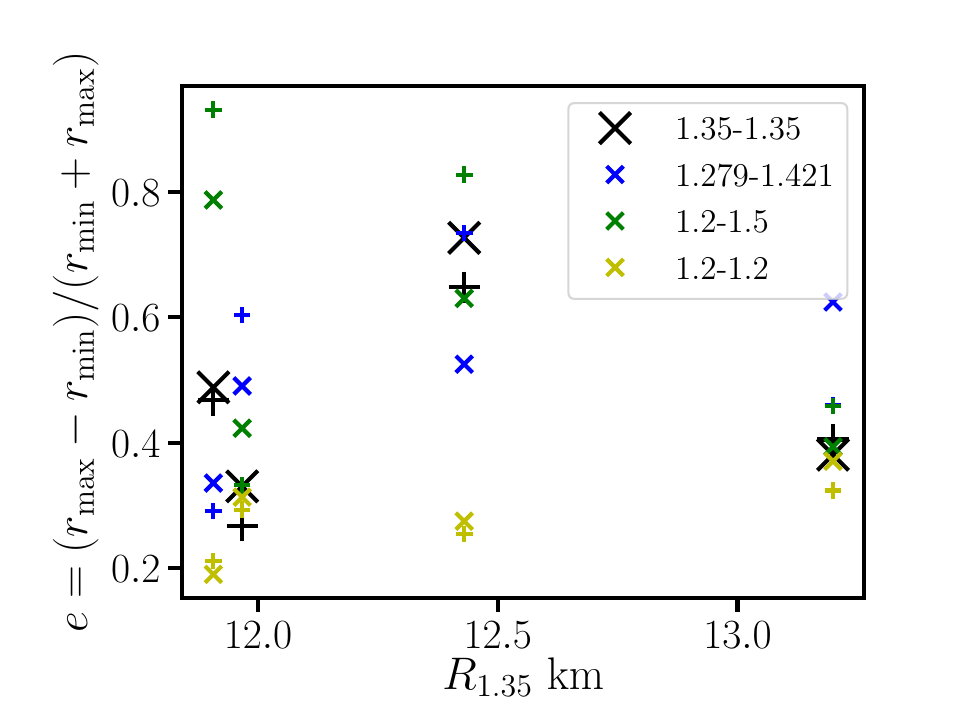}
\caption{Eccentricity of DM motion in the merger remnant as function of NS radius. Symbols have the same meaning as in~Fig.~\ref{fig:fr135}.}
\label{fig:e}
\end{figure}

\begin{figure}
\centering
\includegraphics[width=\columnwidth]{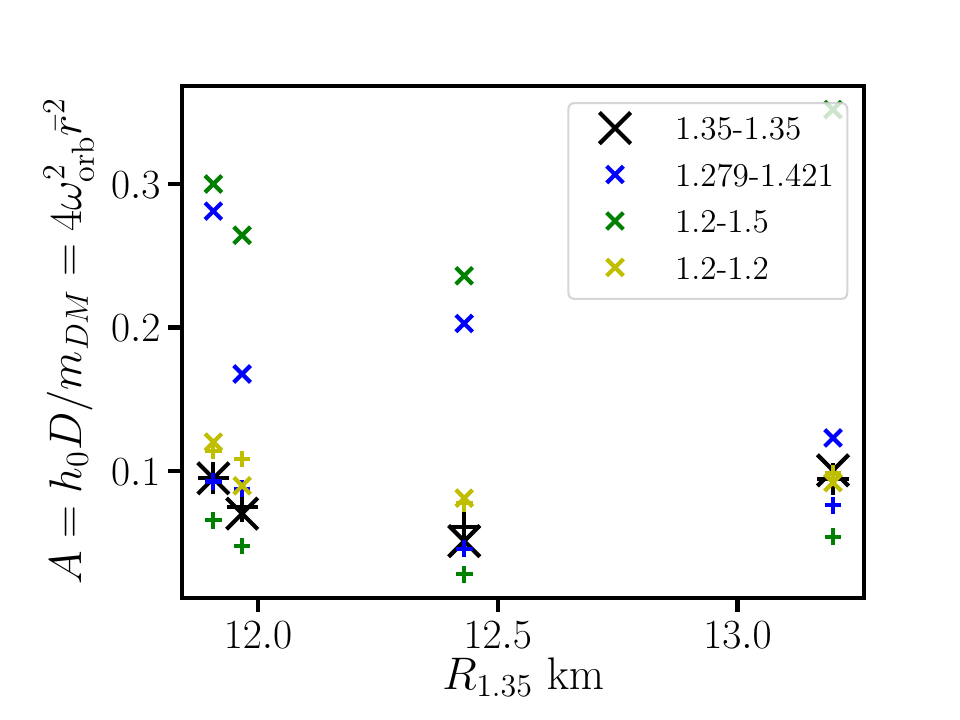}
\caption{Amplitude of GW signal of the DM component assuming circular orbits with averaged orbital radius as function of NS radius. Symbols have the same meaning as in~Fig.~\ref{fig:fr135}.}
\label{fig:amp}
\end{figure}

We extract the orbital parameters at the end of the simulations for different binary setups from the DM trajectories. Figure~\ref{fig:fr135} shows twice the orbital frequency $f_\mathrm{orb}$, which corresponds to the instantaneous GW frequency $f_\mathrm{DM}$ of the GW signal produce by orbiting DM. We display the frequencies as functions of the radius of a nonrotating NS with 1.35~$M_\odot$, which serves as an effective measure for the stiffness of the EoS employed in the respective simulation.

There is a clear EoS dependence of the DM frequencies. The orbital and thus GW frequency of DM increases with the compactness of NSs. This behavior may be expected because softer EoSs lead to higher mean densities of the remnant (see Fig.~13 in~\cite{Bauswein2012a}) and the orbital frequency of test particles is determined by the mass enclosed by the orbit. See discussion of our long-term evolution model below and Ref.~\cite{Horowitz:2019aim}. For uniform density inside the remnant the orbital frequency of a test particle is constant throughout the remnant. In an actual merger remnant the density increases towards the center and thus the orbital frequency is expected to increase for smaller orbits, i.e. as DM particles slowly inspiral inside the remnant.

In Fig.~\ref{fig:fr135} we mark the test particles being originally placed in the different binary components by a cross (from the less massive star) and a plus sign (from the more massive star). The orbital frequencies of DM particles stemming from the less massive component in asymmetric binaries (crosses) are systematically lower. In asymmetric mergers these DM particles are injected on larger orbits (Fig.~\ref{fig:rad}). The central densities in merger remnants in calculations with the same total mass and EoS but different mass ratios are roughly comparable at the end of the simulations. (At the time of merging and during the early remnant evolution asymmetric mergers lead to higher central densities). Circling on larger orbits thus implies that the mean density inside the orbit is lower for asymmetric mergers, which results in a lower frequency. The DM component which originally resided in the more massive progenitor star orbits with a frequency that is roughly comparable to that of the equal-mass merger with the same $M_\mathrm{tot}$. This is in line with the argument just given. Generally, the orbital frequencies of DM stemming from the two different cores of the merging NSs are not too different even for fairly asymmetric binaries.
As for equal-mass binaries with $M_\mathrm{tot}=2.4~M_\odot$, it is not unexpected that they lead to lower orbital frequencies.

We also relate the very same frequency data to other system parameters. Figures~\ref{fig:fpeak} and~\ref{fig:lam} display the frequencies as function of the dominant postmerger frequency $f_\mathrm{peak}$ (see Fig.~\ref{fig:spec}) and the combined tidal deformability $\tilde{\Lambda}$. The combined tidal deformability encodes EoS effects during the GW inspiral and is given by $\tilde{\Lambda}=\frac{16}{13}\frac{(M_1+12 M_2)M_1^4\Lambda_{1}+(M_2+12 M_1)M_2^{4}\Lambda_{2}}{(M_1+M_2)^{5}}$ with the masses $M_{1/2}$ and the tidal deformability $\Lambda_{1/2}=\frac{2}{3}k_2\frac{R_{1/2}}{M_{1/2}}$ of the individual binary components. $k_2$ and $R$ refer to the tidal Love number and the NS radius, both of which are uniquely determined by the EoS~\cite{Hinderer2010,Chatziioannou2020}. For $f_\mathrm{peak}$ as well as for $\tilde{\Lambda}$ we find a relatively tight correlation between the GW parameters describing the emission by the stellar fluid and the GW frequency of DM. (Considering the definition of $\Lambda$ it is clear that the data points for $M_\mathrm{tot}=2.4~M_\odot$ show an offset with respect to those for $M_\mathrm{tot}=2.7~M_\odot$.) These relations may be used to constrain the frequency range where GW emission by DM may occur and thus may be helpful for GW data analysis. The existence of these relations and the similarity between the different plots is expected since these GW parameters of the baryonic matter are known to scale with the NS radius~\cite{Hinderer2010,Bauswein2012}. We thus continue in the following to present results as function of $R_{1.35}$, which allows an easier identification of the data points for a specific EoS, bearing in mind that very similar relations can be expected for $f_\mathrm{peak}$ and $\tilde{\Lambda}$. (We also remark without providing a figure that there is an approximate scaling with the frequency at the maximum of the GW amplitude marking the end of the inspiral.)

The other relevant parameter characterizing the orbital motion of DM particles inside NS merger remnants and thus their GW signal is the orbital separation. In Fig.~\ref{fig:rad} we show the average orbital radius extracted at the end of our simulations. Since orbits are elliptic, we report $\bar{r}=0.5(r_\mathrm{min}+r_\mathrm{max})$ with $r_\mathrm{min}$ and $r_\mathrm{max}$ being the minimum and maximum distance to the center of mass. We do not correct for gauge effects because our estimates of the GW emission are in any case approximate. We find orbits between about 2 and 8~km, which is generally small and as stressed before smaller than the innermost stable circular orbit of a BH forming from the merger remnant. The sizes of the orbit do not show a very strong EoS dependence, which to some extent may be surprising. We only observe a tentative increase of the orbital separation for stiffer EoSs in our sample of models. This seems reasonable since during merging the centers of the original stars, where the DM particles initially reside, should remain more distant from the common center of mass for larger NS radii. Binary mass asymmetry systematically increases the orbits of the DM particles stemming from the less massive progenitor star. This seems reasonable since the less massive star is less bound and thus may inject the DM component on a larger orbit. 

In Fig.~\ref{fig:rad} we indicate the ellipticity of the orbit for the equal-mass mergers by drawing error bars where the lower and upper edges mark $r_\mathrm{min}$ and $r_\mathrm{max}$, respectively. The eccentricity of the different models is summarized in Fig.~\ref{fig:e}, where we do not recognize a particular EoS or binary mass ratio dependence. However, we assume that in particular the eccentricity could be relatively sensitive to numerics. For instance, the data points from the symmetric binary system show a sizeable difference, whereas based on symmetry arguments one would expect the same value. However, we note that the phase difference between the two DM particles at the end of our simulation is still $\pi$ as expected (see lower right panel in Fig.~\ref{fig:snap}). Since the phase evolution is usually hard to resolve in numerical simulations, the agreement with the theoretically expected behavior increases the confidence in our numerical results. 

We finally summarize these simulation results in Fig.~\ref{fig:amp} by plotting the amplitude $A=4 \omega_\mathrm{orb}^2  \bar{r}^2$ with the orbital angular velocity $\omega_\mathrm{orb}=2 \pi f_\mathrm{DM}$. Within a Newtonian model, the amplitude $A=h_0 D/m_\mathrm{DM}$ quantifies the GW amplitude $h_0$ along the polar direction up to the source distance $D$ and the assumed mass $m_\mathrm{DM}$ of a DM particle. Here we employ the quadrupole formula and assume circular orbits with $\bar{r}$ and $m_\mathrm{DM}$ being small compared to the enclosed mass such that the reduced mass is approximated by $m_\mathrm{DM}$ (see e.g.~\cite{Maggiore2008}). The latter approximation at some point breaks down as the DM particles inspiral towards the center of the remnant.

We do not observe a strong impact of the EoS because the effects on orbital frequency and radius approximately cancel. There is a more pronounced impact of the binary mass ratio such that 1.2-1.5~$M_\odot$ binaries lead to a roughly three times higher amplitude compared to equal-mass mergers. Note that only the DM component originating from the less massive component exhibits an increased amplitude. (One may speculate that this increase might be counter-balanced by a less effective accumulation of DM in lighter NSs.)

\section{Detectability}\label{sec:detectability}
\subsection{Long-term model}
The emitted power by the gravitational radiation of DM particles is generally weak in comparison to the emission by the merger itself because the GW amplitude of orbiting DM lumps with masses below $\sim 0.01 M_\odot$ is small. However, as a result the GW back reaction is also weak. If no other strongly dissipative processes are involved (as in our model), the signal may possibly last for a long time and GWs from a DM component in NS mergers may be detectable (cf. estimates in~\cite{Horowitz:2019aim} for isolated stars). As argued above the particularly long emission time distinguishes the DM signal from the emission of baryonic matter, but may pose challenges for GW data analysis. As one of the main results of our study we point out that the gravitational radiation from DM in NS mergers is ultimately limited by the remnant lifetime because the orbits of DM particles are too small to avoid the rapid accretion by a forming BH.

\begin{figure}
\centering
\includegraphics[width=\columnwidth]{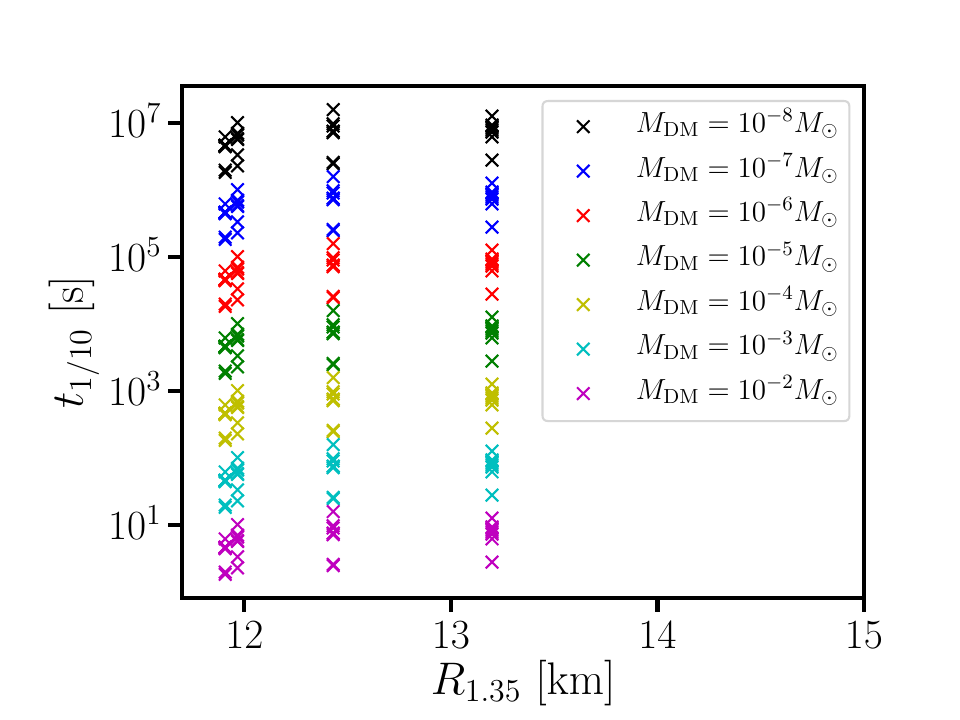}
\caption{Time until the initial GW amplitude of the DM component decayed to 1/10 of its initial value for different simulations discussed in this study. Results are shown as function of the NS radius of the respective EoS used in the simulation. Colors refer to different assumed masses of the DM component.}
\label{fig:t10}
\end{figure}

We now quantify the detectability of GWs from DM particles based on our simulation results. Because of the long signal length (see Fig.~\ref{fig:t10} showing the time scales until the amplitude declined by a factor 10), we cannot compute the GW signal and corresponding signal-to-noise ratio (SNR) self-consistently from our simulation data. Instead we have to employ a model to estimate the long-term evolution and the detectability. Approximations may generally be justified since our model anyway relies on a number of assumptions. In particular, the signal strength is expected to scale with the mass of the DM component, which is unknown and may be even different in different merger events.

We estimate the long-term behavior of the GW signal by assuming Newtonian point-particle dynamics of the DM components under the influence of GW back reaction, which reduces the orbital energy according to the quadrupole formula. We closely follow the model and assumptions in~\cite{Horowitz:2019aim}; see Appendix for details. The approach is fully equivalent to the Newtonian treatment of the binary inspiral phase (which for instance leads to the definition of the chirp mass) except for the difference that the DM particles move inside the stellar remnant. This implies that the gravitational potential differs from that of the classical two-body problem and is determined by the density distribution inside the remnant. To further simplify the estimate, we assume circular orbits\footnote{ We compare the orbital evolution of circular and non-circular orbits for a test particle moving inside a uniform density in Appendix~\ref{app:noncircular} to gauge the impact on our results due to this assumption. We find that the strength of the GW peak may be reduced by a factor two to three for eccentric orbits compared to circular orbits.}
and the DM mass to be small compared to $M_\mathrm{enc}$, which refers to the mass enclosed by a sphere with the orbital radius. We also impose that the baryonic remnant does not undergo any further evolution, i.e. that the density profile remains constant in time\footnote{To some extent we mimic an evolution of the density distribution by assuming a density gradient, which at least allows to roughly estimate the impact.}, and we consider a spherically symmetric distribution of mass in the remnant's center. We neglect any other process which could affect the secular evolution of the orbits on longer time scales like for instance a weak interaction between baryons and DM. Furthermore, we assume that the DM particles are point masses and we do not consider the possibility that the DM components collide during their orbital evolution, which may in fact happen if the DM components have a finite size and move with different orbital frequencies or eccentricity

Under these assumptions an orbiting DM particle generates a GW signal
\begin{equation}\label{eq:h}
h(t)=\frac{4}{D}M_\mathrm{DM}\omega_\mathrm{orb}^2R(t)^2\cos{(2\omega_\mathrm{orb}t)}
\end{equation}
at a distance $D$ with the orbital angular velocity $\omega_\mathrm{orb}$, which may change with time. Here we suppress the dependence on the inclination angle and any initial phase shift, which distinguish the plus and cross polarization. As the orbital radius $R(t)$ evolves, the amplitude slowly decays (although a slight increase of the orbital frequency may occur). The angular velocity is determined by $\omega_\mathrm{orb}^2=\frac{M_\mathrm{enc}(R)}{R^3}$, and thus directly linked to the orbital radius. Hence, a single equation (based on the balance between orbital energy and radiated energy) governing the decay of the orbital radius $R(t)$ determines the whole evolution of the system and especially the GW signal (see Appendix~\ref{app:longterm}).

For uniform density $\rho$, the gravitational potential is such that the orbital frequency remains constant and is proportional to $\sqrt{\rho}$ (cf.~\cite{Horowitz:2019aim}). In this special case, the evolution of the orbital radius is given by an analytic expression, i.e. the GW signal is given by Eq.~\eqref{eq:h} and
\begin{equation}\label{eq:rtana}
    R(t)= \frac{R_0}{\sqrt{2 k R_0^2 t + 1}}.
\end{equation}
with $k=\frac{16}{5} M_\mathrm{DM}\omega_\mathrm{orb}^4$ and the initial orbital radius $R_0=R(t=0)$.
For simplicity we mostly use this expression to estimate the detectability. In the Appendix~\ref{app:longterm} we describe a refined model by which we can assess the effect of a density gradient within the remnant, which leads to a steady increase of the orbital and thus GW frequency. In this case the equation determining the orbital radius and thus the amplitude can only be solved numerically. We sketch the main results of this refined model below and already remark that the SNR is not much different from the value found for the analytic expression. This justifies to estimate the detectability employing the analytic model.

\subsection{Results: detectability}

The SNR as measure for the signal strength and its detectability is given by
\begin{equation}
    \mathrm{SNR}^2=4\int_0^{\infty}df\frac{|\tilde{h}(f)|^2}{S_\mathrm{n}(f)}
\end{equation}
with the Fourier transforms $\tilde{h}(f)$ of the GW amplitude and the noise spectral density $S_\mathrm{n}(f)$, for which we adopt configurations mimicking the projected sensitivity of Advanced Ligo and the Einstein Telescope~\cite{Harry2010,Hild2010}. The Fourier transforms of Eq.~\eqref{eq:h} with Eq.~\eqref{eq:rtana} can be computed analytically if the signal is assumed to start at $t=0$ and to continue until infinity\footnote{We use the expression from \url{https://www.wolframalpha.com}, which becomes somewhat too lengthy to be shown here explicitly.}.
A very long signal length implies the remnant to be stable, which may only be the case for low-mass binaries, and also requires a very long integration time of the GW search, which may be challenging. Hence, the resulting SNR may represent an upper limit. 

\begin{figure}
\centering
\includegraphics[width=\columnwidth]{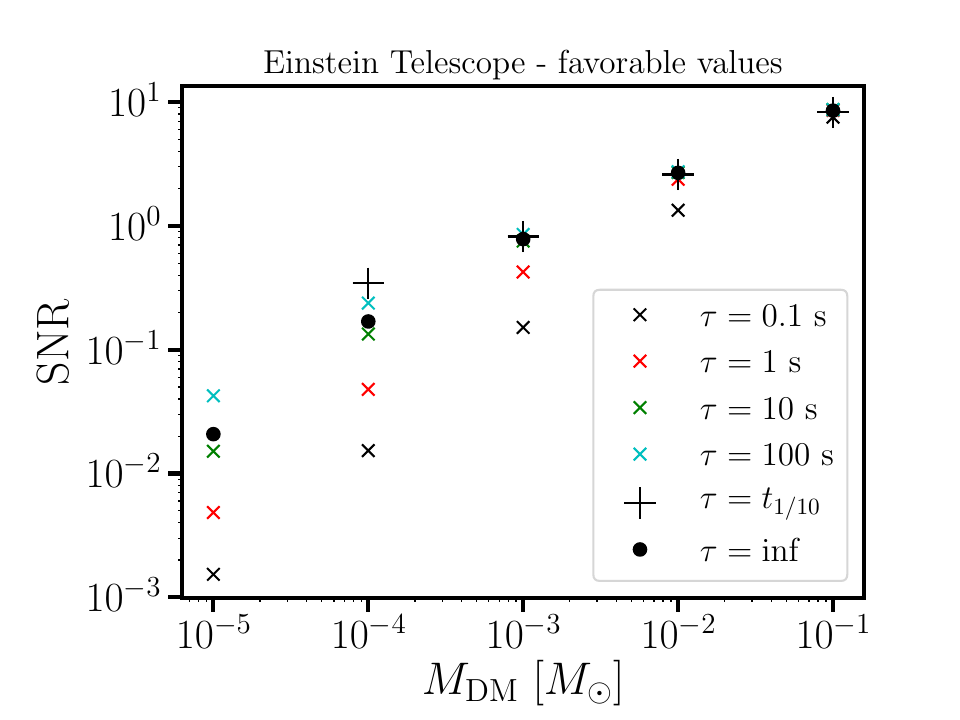}
\caption{Signal-to-noise ratio for the Einstein Telescope as function of DM mass per binary component assuming circular orbits and no frequency evolution. Figure adopts a distance of 20~Mpc (along the polar axis) and generally favorable values for detection, i.e. relatively large orbital radii of 6~km and low orbital frequency of 2~kHz. SNRs are given for different remnant lifetimes $\tau$ or integration time, respectively. Black dots show the SNR for an analytic computation of the Fourier transform of the analytically given signal, corresponding to an infinite lifetime. Black dots thus in principle provide an upper limit on the highest possible SNR. For large $\tau$ the FFT becomes computationally expensive and somewhat inaccurate, which is why some symbols occur at SNR higher than the theoretical limit. $\tau=t_{1/10}$ assumes a lifetime until the initial GW amplitude decayed by a factor 10.}
\label{fig:snret}
\end{figure}

\begin{figure}
\centering
\includegraphics[width=\columnwidth]{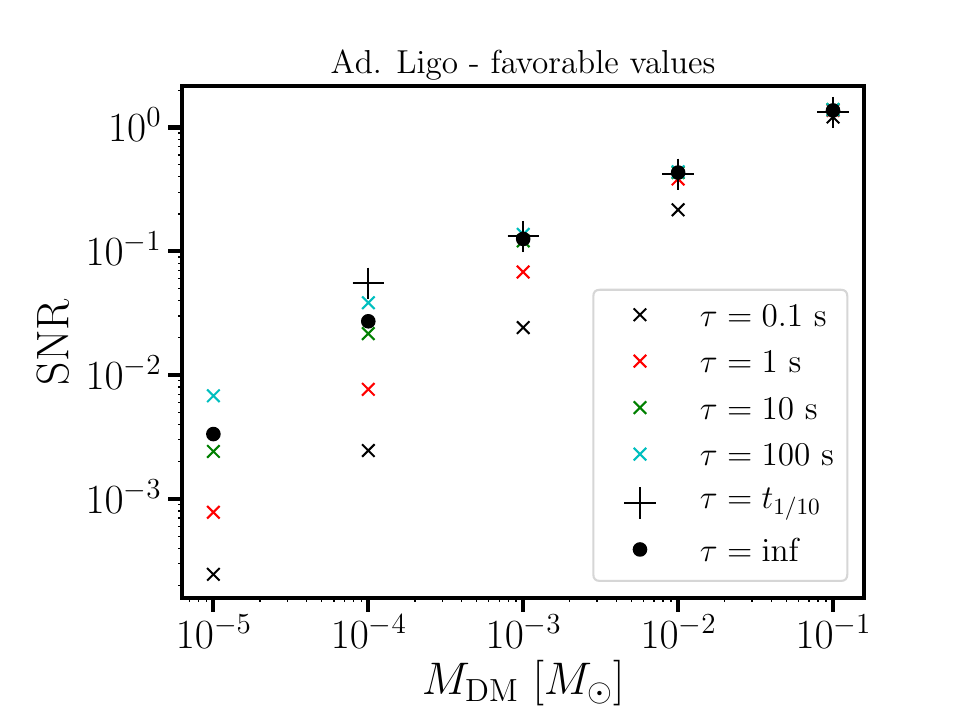}
\caption{Same as Fig.~\ref{fig:snret} but for Advanced Ligo.}
\label{fig:snrad}
\end{figure}

Figures~\ref{fig:snret} and~\ref{fig:snrad} show the SNR for the Einstein Telescope and Advanced Ligo at a polar distance of 20~Mpc as function of the mass $M_\mathrm{DM}$ of the orbiting DM particles (black dots). We assume both binary components to initially host a DM component with $M_\mathrm{DM}$. Here, we compute the SNR for generally favorable but not unrealistic values of the orbital frequency and the initial orbital radius, i.e. adopting $f_\mathrm{orb}=2$~kHz and $R_0=6$~km (see Figs.~\ref{fig:fr135} and~\ref{fig:rad}). One can clearly see that the SNR strongly decreases for lower $M_\mathrm{DM}$. Only for $M_\mathrm{DM}$ as large as $\sim~0.01$ to $0.1~M_\odot$ the SNR becomes significant.

In addition, we consider a finite signal length, which may be limited either by the delayed collapse of the merger remnant or the integration time of the detector. To this end we compute the FFT of Eq.~\eqref{eq:h} for a finite lifetime $\tau$. In Figs.~\ref{fig:snret} and~\ref{fig:snrad} we include the SNRs for lifetimes of 0.1s, 1s, 10s, and 100s (small crosses). Additionally, we compute the SNR for a signal which decays to 1/10 of its initial amplitude (see Fig.~\ref{fig:t10}). As expected the SNR increases for larger $\tau$. The relative difference for different lifetimes increases with smaller $M_\mathrm{DM}$. In the regime of large $M_\mathrm{DM}$ already an integration time of a few seconds may be sufficient, although the general prospects for a detection are not good. We remark that computing the FFT for longer lifetimes becomes computationally expensive (for a desktop machine) because of the many cycles. Note also the slight inconsistencies between the SNR for long but finite lifetime and the infinite lifetime, which also results from the numerical computation of the FFT. (Since this regime is anyway observationally irrelevant we do not attempt to implement further improvements at this point, but caution that our numbers anyway represent only a coarse estimate considering the underlying assumptions.)

\begin{figure}
\centering
\includegraphics[width=\columnwidth]{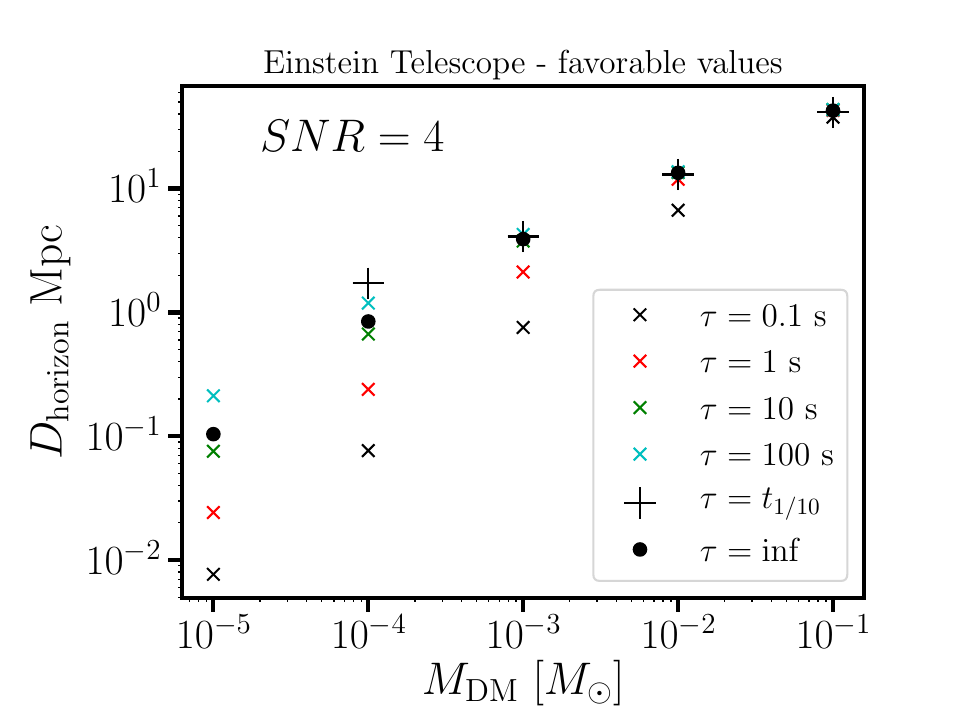}
\caption{Horizon distance for Einstein Telescope as function of DM mass per binary component assuming an SNR of 4 is sufficient for detection.}
\label{fig:horizonet}
\end{figure}

\begin{figure}
\centering
\includegraphics[width=\columnwidth]{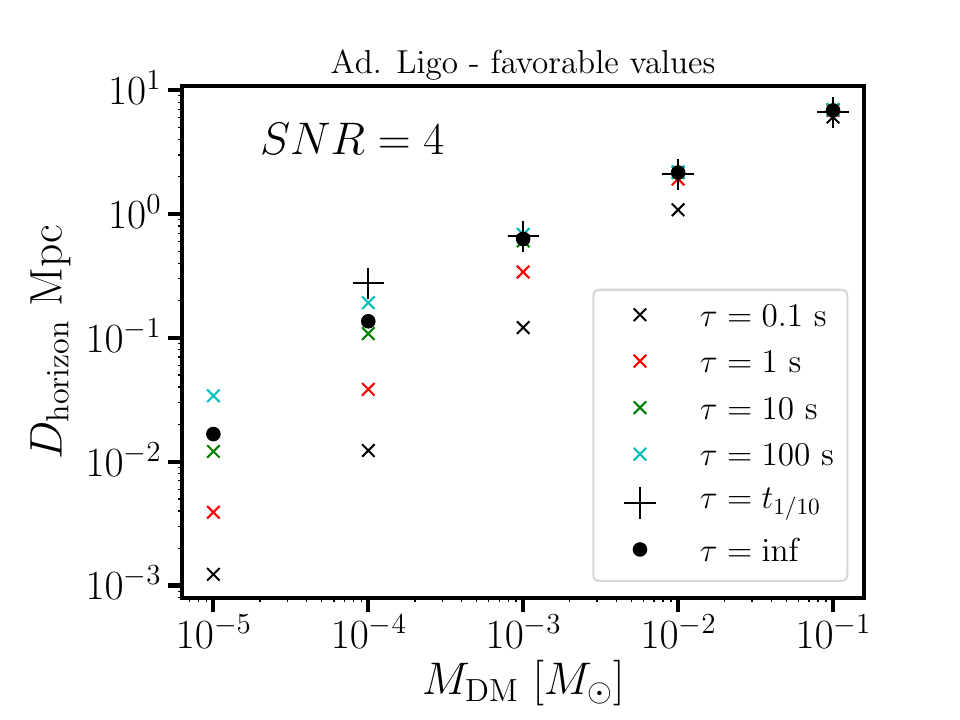}
\caption{Horizon distance for Advanced Ligo as function of DM mass per binary component assuming an SNR of 4 is sufficient for detection.}
\label{fig:horizonad}
\end{figure}

As may already be conceivable from the SNR, the chances for detecting orbiting DM are relatively low and may only succeed under favorable conditions like in particular only large DM masses. We further estimate the detection horizon based on the SNR. Assuming that a SNR of 4 may be sufficient for detection, we display the detector horizons for Advanced Ligo and the Einstein Telescope as function of $M_\mathrm{DM}$ in Figs.~\ref{fig:horizonet} and~\ref{fig:horizonad}. We speculate that a relatively low SNR may suffice because the merger itself may already inform about the potential presence of a signal.

We finally summarize the main conclusions from a refined model of the long-term evolution, which includes a frequency increase because of a density gradient within the NS remnant (see Appendix~\ref{app:longterm} for details). Numerically solving the ordinary differential equation for the evolution of the orbital radius, we obtain the GW amplitude and frequency as function of time. Figure~\ref{fig:amptdd2} in Appendix~\ref{app:longterm} provides an example. We compute the signal for lifetimes of 1s or 10s and adopt different density gradients inside the remnant\footnote{Checking the density distribution towards the end of our simulations the profile is roughly linear with an density increase between a few 10\% and 100\%.} (percentages of density increase from the initial orbital radius to the center of 0\%, 10\% and 100\%). The resulting FFTs are overplotted in the GW spectrum of the NS merger in Fig.~\ref{fig:spec}. Here we consider $M_\mathrm{DM}=0.01~M_\odot$ and refer to Figs.~\ref{fig:snret} and~\ref{fig:snrad} illustrating the dependence on $M_\mathrm{DM}$. The monochromatic signal (for uniform baryon density) exceeds the noise curve of Advanced Ligo for the assume distance of 20~Mpc. However, as already discussed the SNR is relative low since the peak is very narrow. For steeper density gradients the peak becomes flatter and broader and may even drop below the sensitivity curve of the Einstein Telescope. Although the impact on the shape of the peak is significant, comparing signals of the same lifetime the SNR hardly changes if a density gradient is present. Hence, estimates based on the simpler model above are sufficient for our purposes. We note, however, that an evolving frequency increases the complexity of the GW data analysis which may negatively affect the detectability. Similarly, the fact that the orbits are at least initially elliptic may complicate the GW search.

\section{Conclusions}\label{sec:conclusion}
We perform relativistic hydrodynamical simulations of NS mergers and trace the motion of test particles which are assumed to interact only gravitationally with the stellar fluid. By this we model self-interacting DM components which initially reside in the cores of merging NSs. 
A configuration like this may be achieved for strongly self-interacting and fermionic DM that may be produced during supernova explosions or via the anomalous decay of neutrons.
In this study we address whether the GW signal of the orbital motion of the DM component after merging can be detected.

We find that the DM particles remain gravitationally bound and move on elliptic orbits inside the merger remnant. After merging the DM particles decouple from the fluid motion and orbit in the gravitational field of the stellar merger remnant. The orbital radius is relatively small, typically only a few km, where asymmetric binaries yield somewhat larger orbits of up to 8~km. Considering the size of the orbits, we conclude that a forming BH would immediately swallow the DM components and the corresponding GW emission would stop. Hence, the GW signal of DM is limited by the lifetime of the merger remnant. This implies that low-mass binaries may generally provide better chances to detect DM effects in mergers.

We determine 
the frequency of the orbital motion and the corresponding GW signal of DM components in NS mergers from the simulations. The GW frequency is typically in the range between 3 and 5~kHz, depending on the NS EoS and total binary mass, which both determine the merger dynamics and thus the orbital motion of the DM component. We find that the GW frequency of DM roughly correlates with the NS radius, the dominant postmerger oscillation frequency of the stellar fluid and the combined tidal deformability. The latter relations imply that for a given merger event the GW parameters of the stellar fluid provide an estimate in which frequency range the GW emission by DM occurs. The GW frequency of DM is not too sensitive to the binary mass ratio. In fairly asymmetric binaries, the two DM components stemming from the two differently massive progenitor stars have GW frequencies, which differ by several 100~Hz. Here the DM particle from the less massive binary component has a systematically lower frequency, whereas the DM lump from the more massive star results in a frequency very comparable to that of the equal-mass merger with the same total mass. Depending on the long-term evolution of the GW emission, one may expect a splitting of the corresponding peaks in the GW spectrum as suspected in~\cite{Ellis:2017jgp}. 

We extract the amplitude of GWs from DM, which is generally very weak and obviously scales with the assumed mass of DM. Because of its weakness, the signal can last very long (seconds to years) where we assume that any interaction between DM and the stellar fluid is negligible. The signal length clearly distinguishes the DM signature from any GW emission by the stellar fluid which may produce radiation in the same frequency range. Our 
numerical results provide guidance which DM admixture can be excluded if no corresponding signal is detected in the aftermath of a NS merger (at least for the particular class of DM models discussed here).

To estimate the SNR and the detectability of DM in NS mergers, we employ a long-term model accounting for the GW back reaction of the orbital motion. By this we estimate that only DM components as massive as about $0.1~M_\odot$ are detectable by Advanced Ligo ($0.01~M_\odot$ for the Einstein Telescope and comparable third-generation instruments). 
We emphasize that future work should in particular address the requirements for GW data analysis to detect long-lasting GW signals, which may undergo a frequency evolution. Also, our model to estimate the long-term emission of the DM components can be improved by considering relativistic effects, a possible interaction between DM and baryonic matter and a evolution of the density distribution of the merger remnant.
Moreover, we speculate that the GW signals due to the orbiting DM components in eccentric binary NS systems may be louder and may last longer than results shown here assuming quasi-circular orbits. This is because the DM cores can possibly be thrown out to larger radii and thus produce larger GW amplitudes, if they remain gravitationally bound to the merger remnants.
However, the probability of eccentric NS mergers may be much smaller than that of quasi-circular mergers. Also, such type of events may be more difficult to detect because of additional GW data analysis challenges, and eccentric mergers may be less contaminated by DM if they form predominantly in globular clusters with a lower DM content. Future work should also address more extreme binary mass ratios, which may lead to larger orbits and thus stronger GW emission. Similarly, NS-BH mergers should be considered.

We stress that we employ a specific type of DM model with strong self-interaction but a weak coupling to ordinary matter. This is essential to treat DM as test particles. Hence, the described framework and GW signature probes only a certain class of DM models.
Finally, comparing the discovery potential or the sensitivity to a specific DM model with the NS mergers examined here versus the scenario of nearby NSs capturing dark objects~\cite{Horowitz:2019aim} will require further studies including a consistent framework dealing with the production mechanism of these dark objects and their populations, which should be pursued in future work.

\begin{appendix}

\section{Phenomenological origin and the size of the dark object}\label{app:dark_radius}

For DM to form a self-bound compact object, it requires self-interaction in the dark sector.
The DM self-interaction can be either attractive through the exchange of a scalar
boson $\phi$
or repulsive by exchanging a vector boson $V^\mu$ among fermionic DM $\chi$.
The associated interaction Lagrangians are $\mathcal{L}_\phi=g_\phi \phi \bar{\chi}\chi$ and $\mathcal{L}_V=g_V\bar{\chi}\gamma_\mu\chi V^\mu$ where $g_{\phi, V}$ are the coupling constants.
The EoS that describes the dark compact object in the presence of $\mathcal{L}_{\phi,V}$ is given in Ref.~\cite{Gresham:2018rqo}.
The comprehensive discussion on the origin of the phenomenological model
and its implication on the cosmological evolution can be
found in Refs.~\cite{Gresham:2017cvl,Gresham:2017zqi,Gresham:2018anj}.
We proceed theory-agnostically in the following.

Using Table 1 in Ref.~\cite{Gresham:2018rqo}, we can derive scaling relations for the radius $R_D$ and the mass $M_D$ of the dark compact object. 
In terms of strongly attractive self-interaction only,
the relation $R_D\sim R_{\rm max}\times(M_D/M_{\rm max})^{1/3}$, where $M_{\rm max}$ and $R_{\rm max}$ denote the maximum mass of the dark object and the corresponding radius, gives rise to
\begin{equation}\label{eq:DM-scaling}
  R_D\simeq 0.5~{\rm km}
  \left(\frac{C_\phi^2}{10}\right)^{\frac{1}{3}}
  \left(\frac{1{\rm GeV}}{m_\chi}\right)^{\frac{4}{3}}
  \left(\frac{M_D}{0.01{M_\odot}}\right)^{\frac{1}{3}}.
\end{equation}
where $C_\phi=(g_\phi/\sqrt{3\pi^2})(m_\chi/m_\phi)$ is taken to be $\gg 1$, and $m_\chi$ and $m_\phi$ are the DM and mediator masses respectively. 

For strongly repulsive forces with $C_V=(g_V/\sqrt{3 \pi^2})(m_\chi/m_V)\gg 1$, taking $R_D\sim R_{\rm max}$ as an approximation, one obtains
\begin{equation}\label{eq:DM-scaling-rep}
  R_D\simeq 1.0~{\rm km}
  \left(\frac{C_V}{\sqrt{10}}\right)
  \left(\frac{5{\rm GeV}}{m_\chi}\right)^2.
\end{equation}

Eqs.~\eqref{eq:DM-scaling} and \eqref{eq:DM-scaling-rep} clear show that a very compact sub-km dark object with $M_D\lesssim 0.01~M_\odot$ can be obtained for $m_\chi\lesssim 1$~GeV when DM self-interact strongly with an attractive force. 
This is due to mainly the mediator-reduced effective mass of $\chi$ in medium, as detailed in Ref.~\cite{Gresham:2018rqo}.
On the other hand, for strongly-repulsive self-interaction, the dark object will only become sub-km for heavier $m_\chi$.

\section{Long-term model}\label{app:longterm}
We describe a long-term model to estimate the orbital motion of a DM particle moving inside a NS merger remnant closely following the model and assumptions in~\cite{Horowitz:2019aim}. By this we estimate the GW signal of the system. As detailed in the main text, we assume Newtonian point particle dynamics, circular orbits, a fixed background density profile, a small DM mass compared to the mass enclosed by a sphere of the initial orbit $R_0$, spherical symmetry of the mass distribution within $R_0$, and GW back reaction being the only dissipative process.

For simplicity we adopt a linear density profile
\begin{equation}\label{eq:rho}
\begin{split}
    \rho(R)&=\alpha R+\beta \\
           &=(1-\chi)\frac{\rho_0}{R_0} R + \chi \rho_0
\end{split}
\end{equation}
with $\rho_0=\rho(R_0)$ being the density at the initial orbital radius $R_0$. $\chi$ describes by which factor the density increases from $R_0$ to the center $R=0$. Coefficients $\alpha$ and $\beta$ are introduced to somewhat simplify the calculation.

Integration of the density profile Eq.~\eqref{eq:rho} yields the enclosed mass
\begin{equation}
    M(R)=\int_0^R dR 4\pi R^2\rho(R)=\pi\alpha R^4+\frac{4\pi}{3}\beta R^3,
\end{equation}
which determines the orbital angular velocity via
\begin{equation}\label{eq:omega}
    \omega^2=\frac{M(R)}{R^3}=\pi\alpha R +\frac{4\pi}{3}\beta.
\end{equation}
We do not extract density profiles or the enclosed mass from the simulations but for consistency match the initial $\omega_0=\omega(R_0)$ and $R_0$ with the simulation data. This uniquely determine $M(R)$ and $\rho$ for an assumed density gradient, which we express by $\chi$:
\begin{equation}
\begin{split}
    \omega_0^2=\frac{M(R_0)}{R_0^3}&=\pi\alpha R_0+\frac{4\pi}{3}\beta\\
    &=(\pi(1-\chi)+\frac{4\pi}{3}\chi)\rho_0
\end{split}    
\end{equation}
with the definitions of $\alpha$ and $\beta$ from above. This yields
\begin{equation}
    \rho_0=\frac{\omega_0^2}{\pi(1-\chi)+\frac{4\pi}{3}\chi},
\end{equation}
which uniquely determines $\alpha$ and $\beta$ (Eq.~\eqref{eq:rho}), hence the density profile, for a given $\omega_0$ and $R_0$.

With these equations we are equipped to compute the orbital energy. The potential energy after a integration across the density profile Eq.~\eqref{eq:rho} reads (considering only the terms with an explicit dependence on $R$ and ignoring a constant offset)
\begin{equation}
    E_\mathrm{pot}=M_\mathrm{DM}\frac{\pi}{3}R^2(\alpha R+2\beta).
\end{equation}
The kinetic energy is given by
\begin{equation}
    E_\mathrm{kin}=\frac{1}{2}M_\mathrm{DM} R^2 \omega^2= \frac{1}{2} M_\mathrm{DM} R^2 (\pi\alpha R+\frac{4\pi}{3}\beta).
\end{equation}
This sums up to the orbital energy
\begin{equation}
    E_\mathrm{orb}=\pi M_\mathrm{DM} R^2 (\frac{5}{6}\alpha R+\frac{4}{3}\beta).
\end{equation}
The only time-dependent quantity in this equation is the radius and we express the change of the orbital energy as
\begin{equation}
    \frac{d E_\mathrm{orb}}{dt}=\pi M_\mathrm{DM} R \dot{R}(\frac{5}{2}\alpha R+\frac{8}{3}\beta).
\end{equation}
The change of the orbital energy equals the energy carried away by GWs, which for an orbiting point particle is given by
\begin{equation}\label{eq:de1}
    \frac{d E_\mathrm{GW}}{dt}=-\frac{32}{5} \mu^2 R^4 \omega^6,
\end{equation}
where we approximate the reduced mass with $\mu= M_\mathrm{DM}$. 
This results in
\begin{equation}\label{eq:de2}
    \frac{d E_\mathrm{GW}}{dt}=- \frac{32}{5} M_\mathrm{DM}^2 R^4 (\pi\alpha R + \frac{4\pi}{3}\beta)^3,
\end{equation}
which only depends on $R$. Equating Eqs.~\eqref{eq:de1} and~\eqref{eq:de2} finally yields
\begin{equation}
    \frac{dR}{dt}=-\frac{32 \pi^2 M_\mathrm{DM}R^3(\alpha R+\frac{4}{3}\beta)^3}{5(\frac{5}{2}\alpha R+\frac{8}{3}\beta)}.
\end{equation}
This equation can be solved numerically, and its solution $R(t)$ yields the evolution of the GW signal, i.e. its amplitude and frequency through Eqs.~\eqref{eq:h} and~\eqref{eq:omega}. The solution is fully determined by the initial data $\omega_0$, $R_0$ and $\chi$.

\begin{figure}
\centering
\includegraphics[width=\columnwidth]{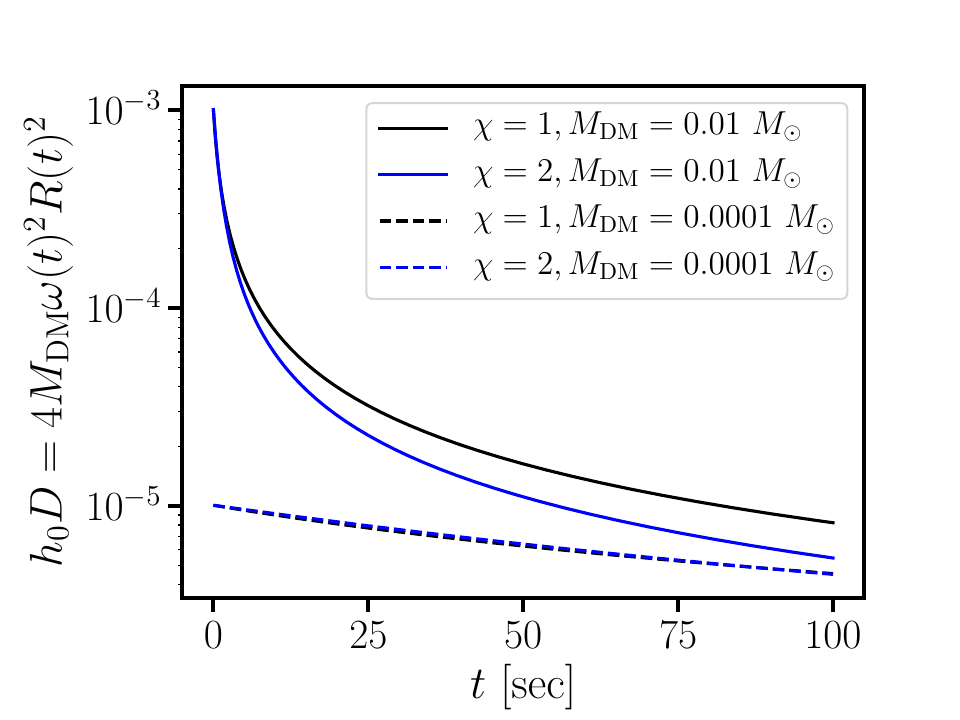}
\caption{GW amplitude as function of time for a 1.35-1.35~$M_\odot$ merger with the DD2 EoS. Initial amplitude is based on simulation data; time evolution is computed by the long-term model described in the appendix making different assumptions about the DM mass and the density gradient inside the remnant (see text). Dashed curves nearly overlap.}
\label{fig:amptdd2}
\end{figure}

For a uniform density $\rho=\rho_0$ we have $\alpha=0$ and $\beta=\rho=\frac{3}{4\pi}\omega^2$ (with $\omega$ being constant). In this case we recover
\begin{equation}
    \frac{dR}{dt}=-\frac{16}{5}M_\mathrm{DM}R^3\omega^4=-k R^3,
\end{equation}
the analytic solution of which is Eq.~\eqref{eq:rtana} with $k$ as given in the main text (see also~\cite{Horowitz:2019aim}).

In Fig.~\ref{fig:amptdd2} we compare the evolution of the amplitude for different density gradients inside the remnant. The impact of the density gradient is relatively weak, which justifies to use the analytic model assuming uniform density to estimate the amplitude. This explains the fact that SNRs are largely unaffected by the frequency evolution because of a density gradient (Fig.~\ref{fig:spec}). For lower DM mass the decay of the amplitude is generally slower. 

\begin{figure*}
\centering
\includegraphics[width=0.9\textwidth]{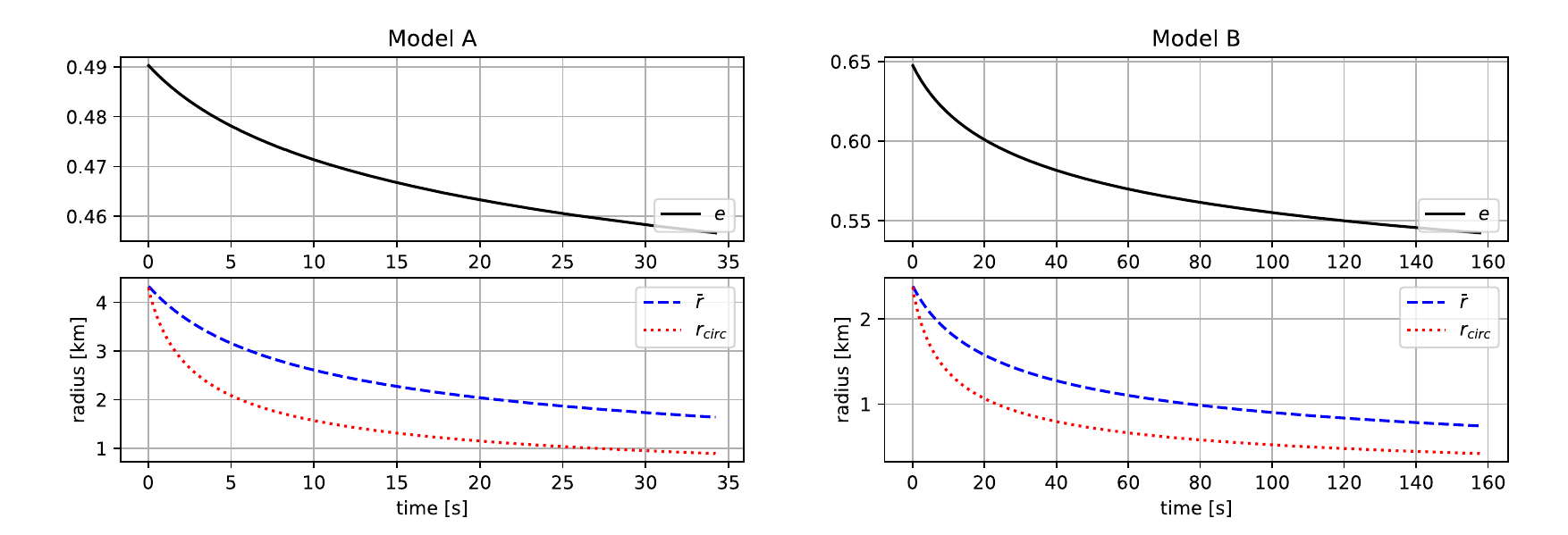} 
\caption{The evolution of $\bar r = 0.5(r_{\rm min}+r_{\rm max})$ (blue dashed) and $e=(r_{\rm max}-r_{\rm min})/(r_{\rm min}+r_{\rm max})$ (black solid) for a test particle moving within a uniform density background with non-circular orbits for four cases listed in Table~\ref{tab:appC-models}. 
Red dotted lines show the corresponding radial evolution of circular orbits $r_{\rm circ}$ assuming initially $r_{\rm circ}=\bar r$.
}
\label{fig:appc-orbit}
\end{figure*}

\section{Effect of non-circular orbits}\label{app:noncircular}
In our long-term model described in Appendix.~\ref{app:longterm}, we assumed immediate circularization of DM particles for simplicity. 
In this Appendix, we compute the evolution of non-circular orbits for a point particle that is immersed in a constant density profile taking the Newtonian limit. 
We compare the obtained orbital evolution and the GW emission to the corresponding quasi-circular cases using two representative cases shown in the main text, which are listed in Table~\ref{tab:appC-models}.

\begingroup
\begin{table}[t]
\begin{ruledtabular}
    \caption{\label{tab:appC-models} Two sets of initial parameters for non-circular orbits extracted at the end of simulations described in the main text. $R_0 = 0.5(r_{\rm min}+r_{\rm max})$ and $e_0=(r_{\rm max}+r_{\rm min})/(r_{\rm max}+r_{\rm min})$.}
    \centering
    \begin{tabular}{ccccc}
        Model & $R_0$ [km] & $e_0$ & EoS & binary masses [$M_\odot$] \\ \hline
        A & 4.331 & 0.49 & SFHX &	1.279--1.421\\ \hline
        B & 2.385 & 0.65 & DD2F & 1.350--1.350\\
    \end{tabular}
\end{ruledtabular}
\end{table}
\endgroup

For non-circular Newtonian orbits, we begin with the Lagrangian
\begin{equation}\label{eq:Lagrangian}
 \mathcal{L}(R,\theta,\dot{R},\dot{\theta}) = \frac{1}{2} \mu(R)(\dot{R}^2 + R^2 \dot{\theta}^2 ) - U(R),
\end{equation}
where $\mu(R)=M_{\rm DM}/M(R)$ is the effective mass of the system in the limit of $M_{\rm DM}\ll M(R)=4\pi\rho R^3/3$, $\theta$ is the polar angle of the point DM particle on the orbital plane relative to the semi-minor axis of the orbit, and $U(R)=-M(R)M_{\rm DM}/R$.

From Eq.~\eqref{eq:Lagrangian}, one can derive the corresponding equation of motion of the system, expressed in terms of $u \equiv 1/R$ (in the test particle limit, i.e., $M_{\rm DM}\ll M(R)$, $\theta$):
\begin{equation}\label{eq:nceom-newton}
\frac{d^2 u}{d \theta^2}  + u =  \frac{1}{L^2}  \left(  \frac{4}{3} \pi \rho M^2_{\rm DM} \right) \frac{1}{u^3}, 
\end{equation}
where $L=M_{\rm DM}R^2\dot{\theta}$ is a constant of motion without GW emission.

Taking $M_{\rm DM}=10^{-3}$~$M_\odot$ and $\rho=5.477\times 10^{14}$~g~cm$^{-3}$, for any given initial conditions of orbital parameters with $R_0=\bar r = 0.5(r_{\rm min}+r_{\rm max})$ and $e_0=e=(r_{\rm max}-r_{\rm min})/(r_{\rm min}+r_{\rm max})$ extracted from numerical simulations described in the main text, we 
first solve Eq.~\eqref{eq:nceom-newton} numerically to obtain the orbit, and use the quadruple formula to estimate the radiated energy and angular momentum by GW for a single orbit.   
We then assume that the system stay on the same orbit for $N_0$ periods until the accumulated energy loss $|\Delta E|>0.01|E_0|$, where $E_0$ is the initial total energy of the orbit. 
Once $|\Delta E|>0.01|E_0|$, we subtract $\Delta E$ and the corresponding $\Delta L$ from the initial $E_0$ and $L_0$ to obtain $E_1=E_0-\Delta E$ and $L_1=L_1-\Delta L$ as the new orbital parameters. 
We repeat this procedure iteratively to obtain the approximate orbital evolution that include the GW emission over a longer period.

\begin{figure*}
\centering
\includegraphics[width=0.9\textwidth]{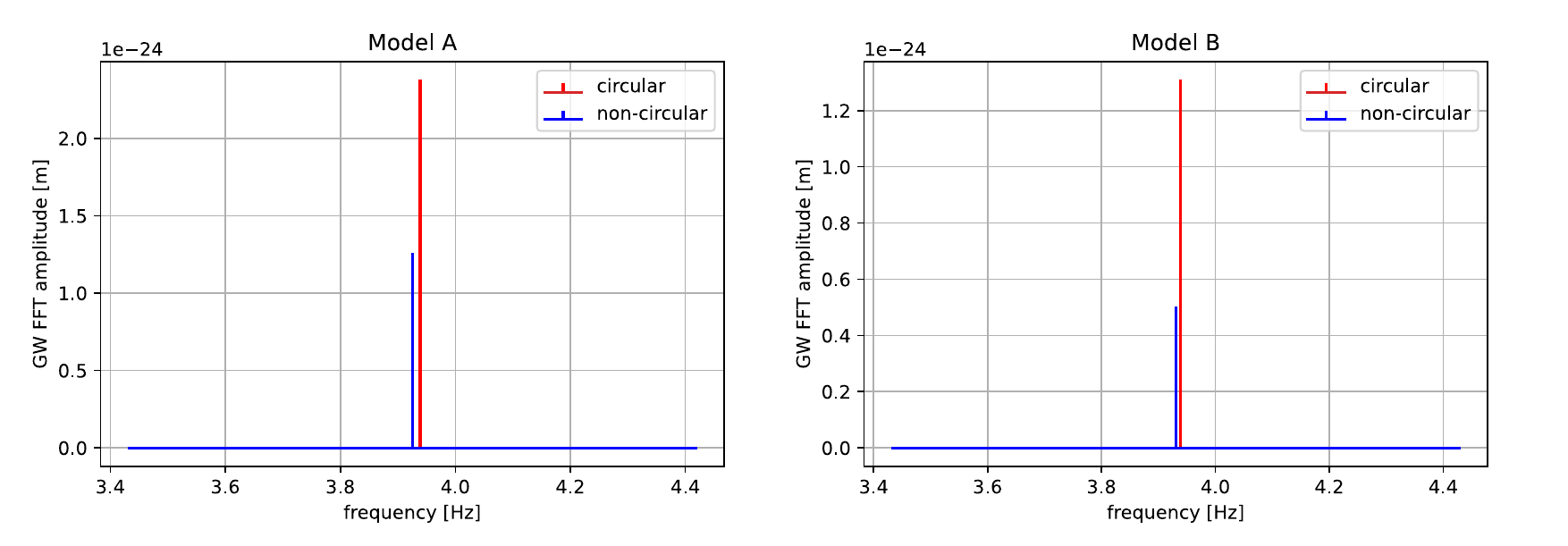} 
\caption{Comparison of the emitted GW amplitudes for the non-circular and circular cases shown in Fig.~\ref{fig:appc-orbit}.}
\label{fig:appc-gw}
\end{figure*}

Fig.~\ref{fig:appc-orbit} shows the evolution of $\bar r$ and $e$ for the two cases listed in Table~\ref{tab:appC-models} as well as the corresponding evolution of the orbital radius $r_{\rm circ}$ assuming circular orbits (see Appendix.~\ref{app:longterm}) for 300 iterations.  
Interestingly, for non-circular cases, $\bar r$ decreases considerably slower than $r_{\rm circ}$, which suggests that the GW emission is less efficient for non-circular orbits than the corresponding circular cases.  
This is mainly because for a test particle moving inside a uniform medium, the enclosed total mass $M\propto R^3$ also varies with radius, differently from the two orbiting point particles. 
For GW emission, the net effect of the smaller enclosed mass dominates over the enhancing factor due to smaller radii, which leads to the observed less efficient GW emission. 
We also note that the eccentricity parameter $e$ decreases very slowly so that we do not expect quick circularization.  

In Fig.~\ref{fig:appc-gw} we show the emitted GW spectra for the same cases as in Fig.~\ref{fig:appc-orbit} over the evolved periods. 
It suggests that the GW amplitude peaks at nearly the same frequency (determined by the constant density of the medium), but the amplitudes for non-circular cases are generally a factor of 2-3 times smaller than the circular cases.

\end{appendix}

\section*{Acknowledgement}
We thank S. Blacker, C. J. Horowitz, G. Lioutas, S. Reddy and F. Roepke for helpful discussions. We thank N.-U. F. Bastian, D. Blaschke, M. Hempel and T. B. Fischer for providing EoS tables. A.B. acknowledges support by the European Research Council (ERC) under the European Union's Horizon 2020 research and innovation programme under grant agreement No. 759253 by Deutsche Forschungsgemeinschaft (DFG, German Research Foundation) - Project-ID 279384907 - SFB 1245 and DFG - Project-ID 138713538 - SFB 881 (``The Milky Way System'', subproject A10) and by the State of Hesse within the Cluster Project ELEMENTS.
G.~G. and M.-R.~W. acknowledge support from the Academia Sinica by Grant No. AS-CDA-109-M11. G.~G. also acknowledges support from the 
National Natural Science Foundation of China under Grant No. 12205258. J.-H. Lien and M.-R.~W. acknowledge support from 
the National Science and Technology Council, Taiwan (NSTC) under Grant No. 109-2112-M-001-004 and  No. 110-2112-M-001-050. 
M.~R.~W also acknowledges support from the NSTC under Grant No. 111-2628-M-001-003-MY4, and Physics Division, National Center for Theoretical Sciences of Taiwan. 
Y.-H.~Lin acknowledges support from the Postdoctoral Scholar Program of the Academia Sinica.

%

\end{document}